\documentclass[twocolumn,english,aps,prapplied,superscriptaddress]{revtex4-1}
\usepackage{bm}
\usepackage{ulem}

\usepackage{amsmath}
\usepackage{graphicx}

\usepackage{xcolor}
\usepackage{soul}
\usepackage{soulpos}

\usepackage{dcolumn}
\usepackage{bm}
\usepackage{subfigure}
\usepackage{mathptmx}

\graphicspath{{figures/}}

\begin{document}

\title{Micromagnetic Modelling of the Heat Assisted Switching Process in\\ high Anisotropy FePt Granular Thin Films}

\author{Lewis J. Atkinson}
\affiliation{Department of Physics, University of York, Heslington, York YO10 5DD, United Kingdom}
\author{Richard F. L. Evans}
\affiliation{Department of Physics, University of York, Heslington, York YO10 5DD, United Kingdom}
\author{Roy W. Chantrell}
\email{roy.chantrell@york.ac.uk}
\affiliation{Department of Physics, University of York, Heslington, York YO10 5DD, United Kingdom}

\begin{abstract}
The dynamic process of assisted magnetic switchins has been simulated to investigate the associated physics. The model uses a Voronoi construction to determine the physical structure of the nano granular thin film recording media; and the Landau-Lifshitz-Bloch (LLB) equation is solved to evolve the magnetic system in time. The reduction of the magnetization is determined over a range of peak system temperatures and for a number of anisotropy values. The results show that the HAMR process is not simply magnetization reversal over a thermally reduced energy barrier. To achieve full magnetization reversal (for all anisotropies investigated) an applied field strength of at least 6kOe is required and the peak system temperature must reach at least the Curie point ($T_{\mathrm{c}}$). When heated to $T_{\mathrm{c}}$ the magnetization associated with each grain is destroyed, which invokes the non-precessional linear reversal mode. Reversing the magnetization through this linear reversal mode is favourable, as the reversal time is two orders of magnitude smaller than that associated with precession. Under these conditions, as the temperature decreases to ambient, the magnetization recovers in the direction of the applied field, completing the reversal process. Also the model produces results which are consistent with the concept of thermal writability; when heating the media to $T_{\mathrm{c}}$, the smaller grains require a larger field strength to reverse the magnetization.
\end{abstract}

\maketitle

\section{Introduction}

The increase in magnetic storage (areal) density over the past 60 years has been truly remarkable and is comparable with the much quoted increase in transistor device density - Moore's Law. The increase in areal density has in the main resulted from an adherence to geometric scaling\cite{McDaniel,Wood} as the characteristic dimensions of the bits have been reduced over successive time intervals; although other factors have impacted on the increase in areal density, such as the change from longitudinal to perpendicular recording. Currently information is recorded on a length scale of nanometers. At such length scales the thermal stability of the magnetic components within the recording media is an increasingly significant problem, due to the onset of superparamagnetic effects \cite{Bean}. For a usable storage lifetime the following inequality must be satisfied.

\begin{equation}\label{eqn:no}
    n_{o}=\frac{K_{u}V}{k_{B}T} > 60
\end{equation}

Where $n_0$ is the thermal stability factor. The historic solution to maintaining this inequality, when reducing the size of the magnetic components, has been to use materials with a larger bulk perpendicular magnetocrystalline anisotropy (K$_{u}$), which are thermally stable at a smaller volume (V)\cite{McDaniel}. However, the maximum output field of the inductive write head is determined by the material limit of the highest saturation magnetization found in magnetic solids, $\mu_{0}M_{s} \sim $ 2.5 T\cite{McDaniel,Granz}. Therefore K$_{u}$ cannot be scaled up to the desired level to maintain $n_{o}$, since the write field required to reverse the magnetization will soon exceed the maximum output field of the inductive write head. Therefore the maximum value of K$_{u}$ is limited by the capability of write head technologies\cite{McDaniel}.

A possible solution could use the fact that the anisotropy energy of a ferromagnetic material decreases with temperature, tending to zero as the temperature approaches the Curie point ($T_{\mathrm{c}}$)\cite{McDaniel,Heinonen}. Therefore $n_{o}$ can be varied from a condition of media stability at ambient temperature, to approximately zero at temperatures approaching $T_{\mathrm{c}}$. Thus the temperature of the storage material can be raised locally and momentarily, solving the problems of writability via the low anisotropy state, with thermal stability being regained on cooling back to ambient temperature. This technique would allow for increasingly small grains to be used in the storage medium, leading to an increase of the areal density\cite{Heinonen}. One technology which aims to use this approach is heat assisted magnetic recording (HAMR). 

The L1$_{o}$ ordered FePt, granular thin film is one of the most promising candidates for HAMR media due to its high bulk perpendicular magnetocrystalline anisotropy constant K$_{u}$(0) = 9.0$\times$10$^{7}$ ergs/cc\cite{Granz} and a relatively low Curie temperature. HAMR is presently moving toward practical realization; recent work\cite{Ju} having achieved a demonstrated areal density of 1.4 Tbit/sq in, for the first time overtaking the densities achieved in conventional perpendicular recording. Despite this, the physical reliability of the HAMR process for realistic media is poorly understood, particularly considering the effects of grain size and its distribution and the importance of the anisotropy and intergranular exchange on the reversal process.

In this paper we present a study of the magnetization dynamics of an FePt granular thin film during the HAMR process. An important factor in the model is the use of the Landau-Lifshitz-Bloch (LLB) equation to model the magnetization dynamics of the individual FePt grains within the thin film. The LLB equation has been used in place of the historically used LLG equation, as it better describes the magnetization dynamics at elevated temperatures, close to $T_{\mathrm{c}}$\cite{Chubykalo}. Importantly, the LLB equation allows magnetization reversal via the linear reversal mechanism\cite{Kazantseva,Barker} which is important at temperatures close to $T_{\mathrm{c}}$. The reversal process of the thin films magnetization and individual grains are investigated over a range of values for $K_{v}$ and maximum temperatures.

We show that the HAMR process is not as simple as magnetization reversal within a reduced anisotropy regime. Reversal of the thin film magnetization is only achieved where the temperature is raised to $T_{\mathrm{c}}$ and the applied field is sufficient to overcome the thermal excitation introduced via the heat assist.

\section{Theoretical Model}

To investigate the HAMR process a model has been developed, which replicates the physical structure of the granular recording medium, correctly describes the magnetization dynamics over a broad temperature range and includes any granular interactions.

\subsection{Magnetization Dynamics: The Landau Lifshitz Bloch Equation}

The dynamics of the magnetization of each individual grains are modeled as single macrospins, using the Landau-Lifshitz-Bloch (LLB) equation. The LLB equation is capable of modeling the magnetization \textbf{m}$_{i}$ of ferromagnetic nano particles on the sub pico-second time scale and at temperatures up to and above $T_{\mathrm{c}}$\cite{Garanin,Garanin_2}, and is here given in the stochastic form proposed by Evans et.al.\cite{Evans}, as;

\begin{equation}\label{eqn:LLB}
\begin{split}
    \dot{\textbf{m}}_{i}=-|\gamma|[\textbf{m}_{i}\times \textbf{H}_{\mathrm{eff}}^{i}]+\frac{|\gamma|\alpha_{\|}}{m_{i}^{2}}(\textbf{m}_{i}\cdot\textbf{H}_{\mathrm{eff}}^{i})\textbf{m}_{i}\\\\
    -\frac{|\gamma|\alpha_{\bot}}{m_{i}^{2}}
    \textbf{m}_{i}\times[\textbf{m}_{i}\times (\textbf{H}_{\mathrm{eff}}^{i}+\eta_{\bot})]]+\eta_{\|}
\end{split}
\end{equation}

\bigskip

\noindent where $\alpha_{\|}=\lambda 2T/3T_{c}$ is the longitudinal damping parameter, $\alpha_{\bot}=\lambda(1-T/3Tc)$ is the perpendicular damping parameter, $\gamma$ is the gyromagnetic ratio and $\lambda$ is the intrinsic damping constant which couples the spin system to the heat bath at the atomistic level. Above $T_{\mathrm{c}}$ $\alpha_{\|}=\alpha_{\bot}=\lambda 2T/3T_{c}$. The effects of a non-zero temperature are introduced via a fluctuating magnetic field whose properties are determined by;

\begin{eqnarray}
\langle\eta^{\mu}_i\rangle&=&0, \nonumber\\
\langle\eta_{i}^{\perp}(0)\eta_{j}^{\perp}(t)\rangle&=&\frac{2k_{\rm B}T (\alpha_{\perp}-\alpha_{||})}{|\gamma|M_{\rm s}^{0}V\alpha_{\perp}^2}\delta_{ij}\delta(t), \nonumber\\
\langle\eta_{i}^{||}(0)\eta_{j}^{||}(t)\rangle&=&\frac{2|\gamma|k_{\rm B}T \alpha_{||}}{M_{\rm s}^{0}V}\delta_{ij}\delta(t)
,\nonumber\\
\langle\eta_{i}^{||}\eta_{j}^{\perp}\rangle&=&0,
\label{eqn:randField}
\end{eqnarray}

\bigskip

\noindent and $\textbf{H}_{\mathrm{eff}}$ is the sum of the applied, anisotropy, Zeeman and any interaction fields, which here includes the dipole approximation to the magnetostatic field and the intergranular exchange interaction.

The dynamics of the individual grains are calculated using the Heun scheme to numerically integrate the LLB equation. The parameters for the LLB equation (M$_{s}$(T) and the longitudinal and transverse susceptibilities) are that of FePt and are determined from atomistic calculations using a multi-scale approach\cite{Kazantseva_2}. The Curie temperature is set to 660K and damping parameter $\lambda$ is set to 0.1.

\subsection{Thin Film Structure: The Voronoi Construction}

The model uses a Voronoi construction to produce the granular microstructure of the thin film. The Voronoi construction produces a physically realistic picture of the film, including a grain size dispersion and some microstructural disorder. The model generates three dimensional grains 3 - 10nm in diameter and height, each assumed magnetically to behave as a single domain. The easy axis for each of the grains is aligned with the z-axis. The Voronoi construction produces a film with a dispersion of grain size which is approximately log normal. The standard deviation $\sigma_v$ can be varied by changing the degree of randomness of the seed points for the Voronoi construction. Here we studied a system of 7558 grains with period boundary conditions and a $\sigma_v\sim 0.35$. The local field acting on a grain is calculated as follows;

\begin{equation}\label{eqn:Heff}
\begin{split}
    \textbf{H}_{eff}^{i}=\textbf{H}_{applied}+\sum_{i\ne j}\frac{\mu_{j}}{r_{ij}^{3}}(3(\hat{r}_{ij}\cdot\hat{\mu})\hat{r}_{ij}-\hat{\mu}_{j}) \\\\
    +\textbf{H}_{e}^{i}-\frac{(m_{x}^{i}e_{x}+m_{y}^{i}e_{y})}{\chi_{\perp}} \\\\
    + \left\{ \begin{array}{lr}
    \ \ +\ \frac{1}{\tilde{\chi_{\|}}}\left(1-\frac{m_{i}^{2}}{m_{e}^{2}}\right)\textbf{m}_{i} & T \ \lesssim T_{c}\\\\
    \ \ -\ \frac{1}{\tilde{\chi_{\|}}}\left(1+\frac{3}{5}\frac{T_{c}}{T-T_{c}}m_{i}^{2}\right)\textbf{m}_{i} & \ \ \ T \ \gtrsim T_{c}
    \end{array}
\right.
\end{split}
\end{equation}

\bigskip

\noindent where $\textbf{H}_{applied}$ is the (spatially uniform) external applied field which is set in the negative z direction throughout. The second term on the RHS of Eqn$~\ref{eqn:Heff}$ is the dipole approximation to the magnetostatic term, which is calculated as a direct pairwise summation within a cutoff radius of 8 grain diameters, with long-range effects introduced via a mean field. The remaining terms in Eqn.$~\ref{eqn:Heff}$ are the contributions from the  intergranular exchange, the anisotropy and the intra-granular exchange energy resulting from thermally induced longitudinal fluctuations of the magnetization.

The intergranular exchange interaction is formulated as in the work of Peng et.al.\cite{Peng} under the assumption that the exchange energy is proportional to the contact area between neighboring grains. The approach also allows for a dispersion in the exchange J$_{ij}$ due to variations in grain boundary thickness and composition. In terms of reduced parameters (relative to the median values L$_{m}$, A$_{m}$ and J$_{m}$).

\begin{equation}\label{eqn:exch1}
    \textbf{H}_{e}^{i}=H_{exch}\left(\frac{J_{ij}}{J_{m}}\right)\left(\frac{L_{ij}}{L_{m}}\right)\left(\frac{A_{m}}{A_{i}}\right)\hat{m}_{j},
\end{equation}

\bigskip

\noindent where L$_{ij}$ is the contact length between grains i and j, A$_{i}$ is the area of grain i in the plane of the thin film. In practice H$_{exch}$ is set by the requirement that the average exchange at saturation has a certain value H$_{exch}^{sat}$, that is;

\begin{equation}\label{eqn:exch2}
    H_{exch}^{sat}=H_{exch}\sum_{i}^{N}\sum_{j\in n.n}\left(\frac{J_{ij}}{J_{m}}\right)\left(\frac{L_{ij}}{L_{m}}\right)\left(\frac{A_{m}}{A_{i}}\right),
\end{equation}
It is known that the intra-granular exchange decreases with increasing temperature, the effective field in a mean field approximation scaling with the magnetization. Here we make a similar assumption. The exchange fields given during the later calculations refer to the 0K value of exchange field which is scaled appropriately in the model to represent the temperature reduction of the exchange.
\bigskip

\section{Results}

Here we are not aiming to simulate the detailed HAMR recording process; rather the intention is to study the physics of HAMR as a field cooled magnetization process, which casts light not only on the dynamics of HAMR but also on the thermodynamic aspects limiting the recording densities in HAMR. Consequently, similarly to the field, the temperature of the system is assumed spatially uniform. The anisotropy of the individual grains are modeled as percentage of the bulk magnetocrystalline anisotropy constant, K$_{u}$, of L1$_{o}$ ordered FePt. This models a system in which each grain is partially L1$_{o}$ ordered, giving an anisotropy equal to a percentage of L1$_{0}$ ordered FePt. Each grain within the thin film has an equal K$_{u}$.

\bigskip

\begin{figure}
  \includegraphics[width=8.0cm]{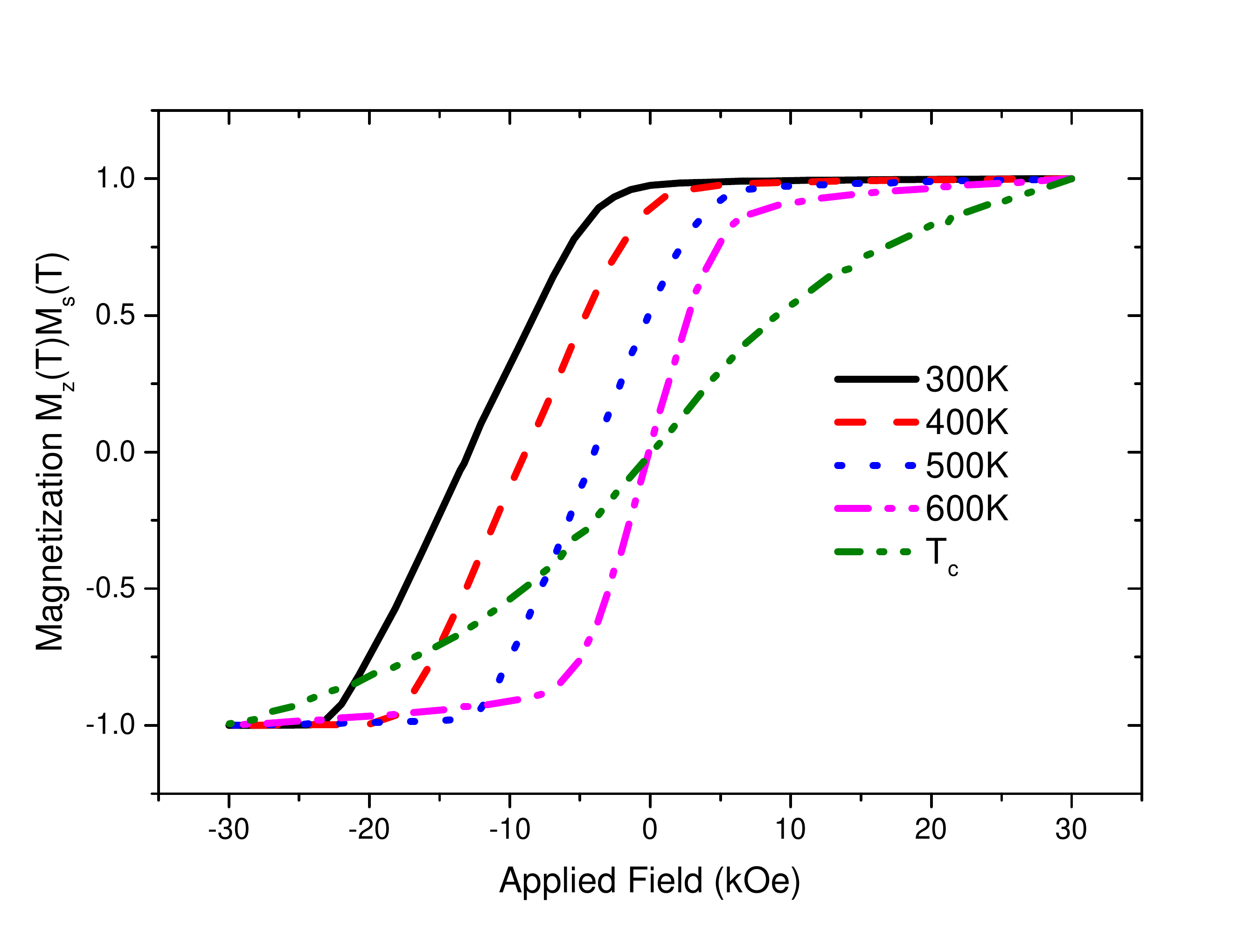}
  \caption{(Color online) Hysteresis curves generated with a thin film comprised of grains with an average diameter of 6.0nm with an anisotropy of 20\% L1$_{o}$ FePt case. These hysteresis curves show the effect of increased temperature.}
  \label{fig:HYST20L1o4000OeExch}
\end{figure}

The initial study consists of simulations using the minimum anisotropy required to ensure the desired stability of the magnetization when applying equation~\ref{eqn:no} to a system comprised of grains with an average diameter of 6.0nm, this equates to an anisotropy value of 1.8x10$^{7} $ergs/cc, which is approximately 20\% of the anisotropy of fully L1$_{0}$ ordered FePt. To observe the effect of increased temperature on the reversal process, quasi-static hysteresis curves, equivalent to VSM measurements, were generated using this system, the results being given in Fig.~\ref{fig:HYST20L1o4000OeExch}. The quasi-static hysteresis curves
were generated with a field sweep rate below which the observed hysteresis curves did not change - this was found to be approximately 100Oe/ns after a equilibration period, in the high field environment, of 1.0ns. As expected the field required to reverse the magnetization is reduced as the temperature (T) is increased.

\begin{figure}
  \includegraphics[width=8.0cm]{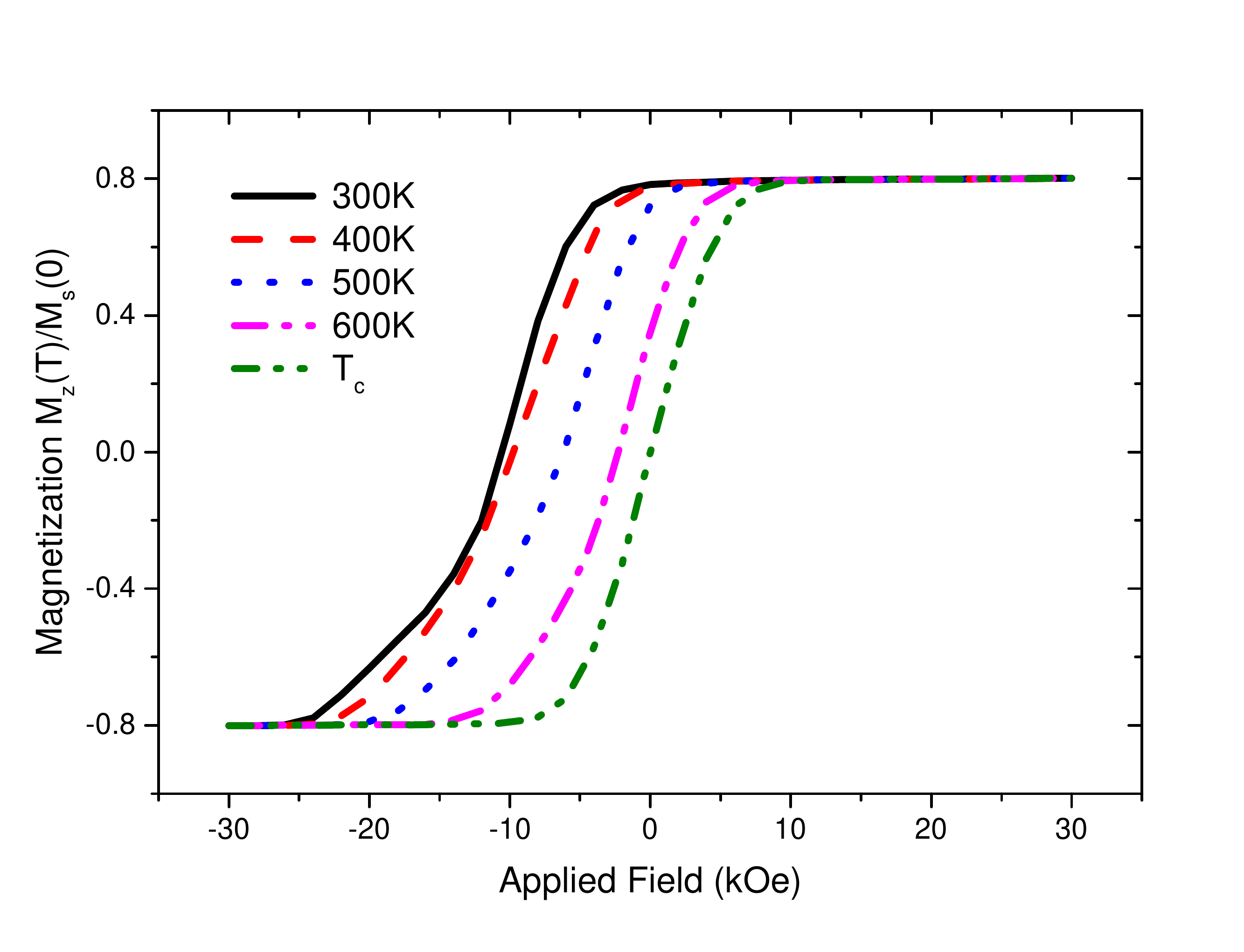}
  \caption{(Color online) Hysteresis curves, using the alternate measurement technique developed by Liu et.al., for the 20\% L1$_{o}$ case, showing the effect of increasing the maximum temperature during the HAMR process.}
  \label{fig:HYST20L1o-FUDAN}
\end{figure}

At ambient temperature (300K) the system requires an applied field in the region of 25kOe to reverse approximately 95\% of the magnetization; clearly the medium cannot be switched using conventional methods. These results are in agreement with the conventional picture of HAMR as magnetization reversal over a thermally reduced energy barrier, although this does not show the full complexity of the HAMR process; as the maximum temperature reached approaches $T_{\mathrm{c}}$, the field required to reverse the magnetization is seen to increase and is of the same order as observed at ambient temperature. This is a reflection of the 'thermal writeability' introduced by Evans et. al.\cite{Evans_2,Richter}, which quantifies the tendency of backswitching of magnetization at elevated temperatures.

Fig. \ref{fig:HYST20L1o4000OeExch} presents measurements over a relatively long timescale, however HAMR differs in nature from these hysteresis experiments as the temperature is raised only momentarily and the reversal process occurs during the subsequent cooling. This process is better investigated using an alternative measurement technique, first performed by Liu et al\cite{Liu}, where the magnetization is observed using a photomagnetic synchronised TR-MOKE technique, which resets the magnetic state to the initial conditions between each pump-probe experiment. Therefore the photomagnetic synchronised TR-MOKE technique determines only the effect of the current laser pump and applied field, independent of the measurement history. It is essentially a single-shot technique with a characteristic timescale similar to that of the HAMR process. We have replicated this technique to better show the effect of the HAMR process. The temperature increase associated with the laser pulse is modeled as a step change from ambient temperature (300K) to the maximum temperature. The temperature then returns to ambient following a Gaussian form in a characteristic time, set to 0.5ns throughout the investigation unless otherwise stated. With this cooling rate the system temperature returns to ambient in $\sim$ 1.2ns. All temperatures stated in the results section are given as the maximum temperature during the HAMR simulation. The decrease in the field required to reverse the magnetization can clearly be seen, in Fig.~\ref{fig:HYST20L1o-FUDAN}, reducing from approximately 25kOe according the standard hysteresis loop of Fig. \ref{fig:HYST20L1o4000OeExch} to 6kOe using the. Using this technique also indicates a reduction in the effect of the intergranular interactions with increasing T, as follows. At low temperatures the curves exhibit a change of gradient in the high negative field. This effect is usually interpreted as arising from the effects of the exchange and magnetostatic interactions, which drives the system into low energy states with low magnetization which are relatively stable and slow the approach to negative saturation. It can be seen in Fig.~\ref{fig:HYST20L1o-FUDAN} that this effect is reduced with increasing temperature, and is not present at the highest temperatures investigated.

\begin{figure}
  \includegraphics[width=8.0cm]{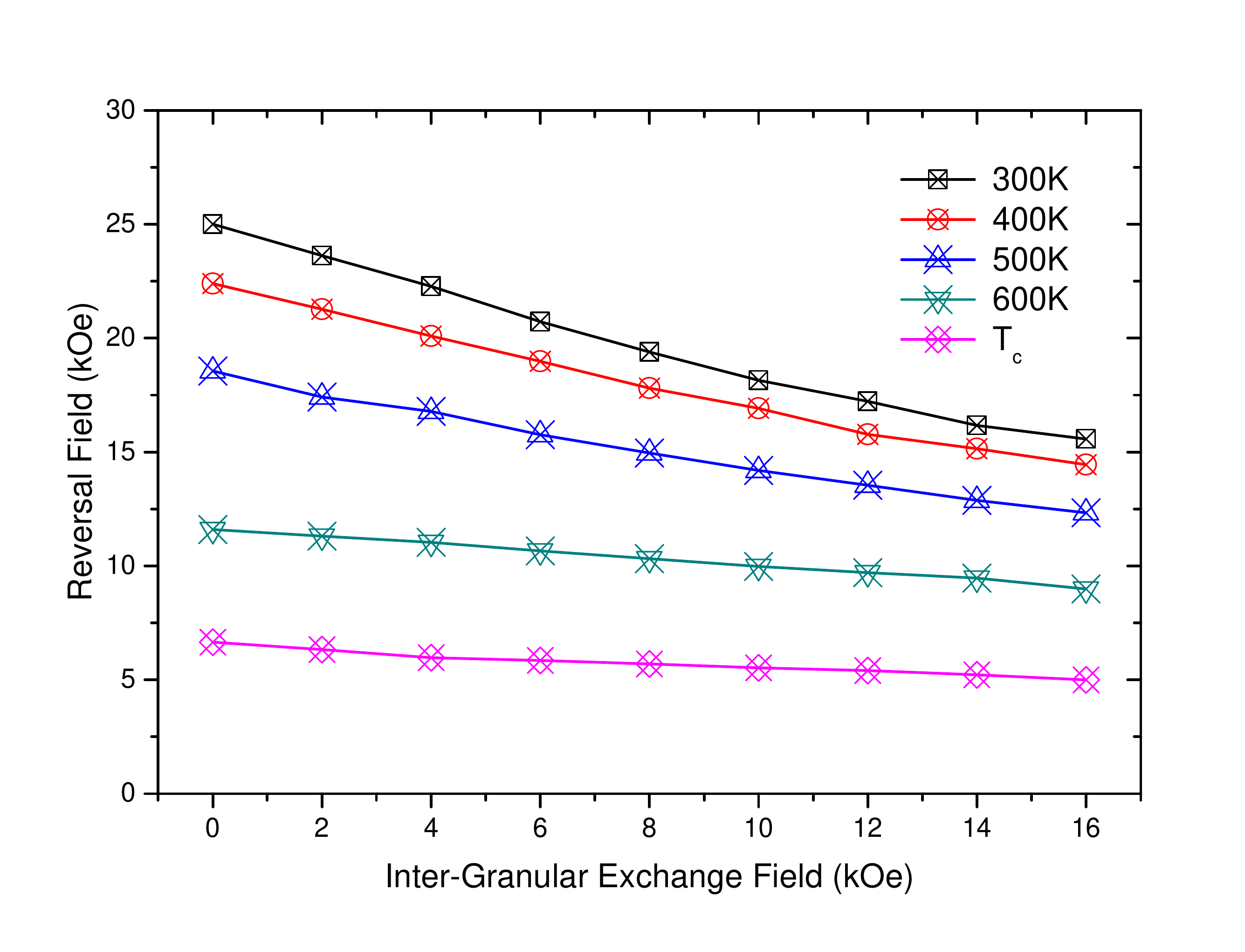}
  \caption{(Color online) The effect on the reversal field as a function of the exchange field and the maximum system temperature during the HAMR process. Note the reduced effect of the exchange field at higher maximum temperatures.}
  \label{fig:EXCHReversalField}
\end{figure}

\begin{figure}
  \includegraphics[width=8.0cm]{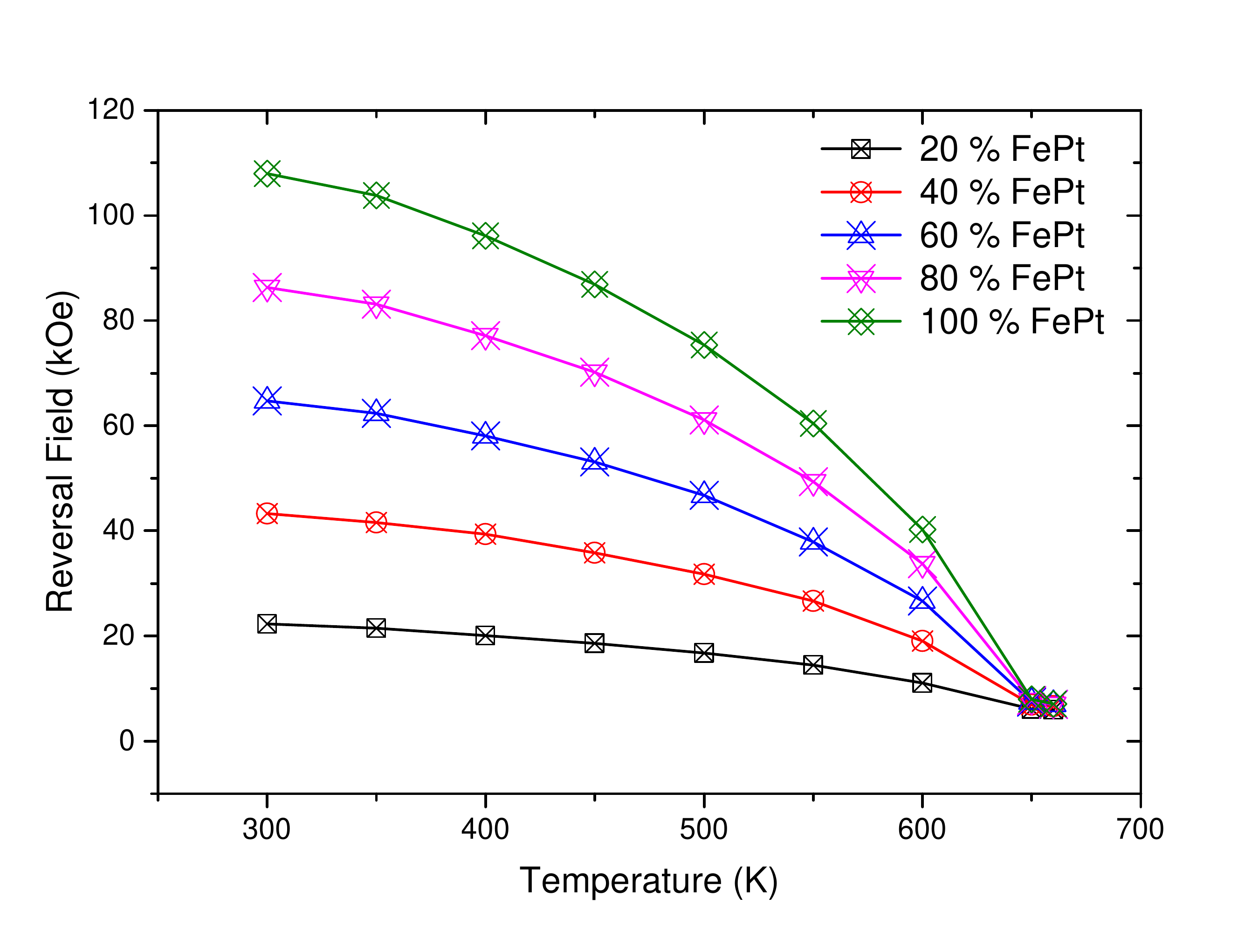}
  \caption{(Color online) The effect on the reversal field as a function of the magnetocrystalline anisotropy (K$_{u}$) and as the maximum system temperature during the HAMR process. At ambient temperature the field required to reverse the magnetization increases approximately linearly with increased K$_{u}$. Increasing the temperature to T$_{c}$ during the HAMR process lowers the reversal to a similar level, 6kOe, over the complete range of anisotropies investigated.}
  \label{fig:KvReversalField}
\end{figure}

The reduction of the effects of intergranular interactions at elevated temperatures is further demonstrated in Fig.~\ref{fig:EXCHReversalField}. Here the reversal field (defined as the field required to reverse 95\% of the thin film magnetization) is given as a function of the intergranular exchange (IGE) field for a range of temperatures from ambient to $T_{\mathrm{c}}$. The variation in the reversal field, due to increasing the IGE field, becomes less pronounced with increasing temperature, suggesting that the effect of the IGE field on the magnetization reversal process decreases with increasing temperature. This suggests that the intergranular interactions are not a significant contribution to the high temperature magnetization processes involved in HAMR.

To further investigate the limits of the HAMR process the effects of K$_{u}$ were further studied, Fig.~\ref{fig:KvReversalField}. At ambient temperature the reversal field is increased approximately linearly with increasing K$_{u}$, which neatly demonstrates the problem facing hard disk technologies. However, with a temperature rise to T$_{c}$ the reversal field is reduced to a similar level, for all values of K$_{u}$ investigated, approximately 6kOe.

This study of the hysteretic behavior confirms that increasing the temperature, as in the HAMR process, significantly reduces the field required to reverse the magnetization when compared with the ambient temperature. Also the data have demonstrated that increasing the temperature reduces significantly the effects of the intergranular interactions and anisotropy field.

\subsection{Dynamics of the HAMR process}

HAMR is essentially a Field-Cooled Magnetization (FCM) process, albeit at extreme rates of temperature reduction and with a spatially and temporally varying magnetic field. This best characterizes the high-temperature switching of the magnetization and its eventual freezing into a stable state at a lower temperature.

Essentially the HAMR process is not simply magnetization reversal over a thermally reduced energy barrier. Evans and Fan~\cite{Evans_3} have shown that, although thermally activated switching below the Curie temperature can result in switching, the timescale over which the temperature must remain constant is prohibitively long. It was suggested in Ref.~\onlinecite{Evans_3} that reliable switching requires heating above $T_{\mathrm{c}}$ so as to activate the linear reversal mode (to be discussed in detail later) during cooling and an applied field strength sufficient to overcome the effects of thermal writability. Here we investigate these thermodynamic aspects of the HAMR process, with emphasis on the effects of the dispersion of grain volume.

\begin{figure}
  \includegraphics[width=8.0cm]{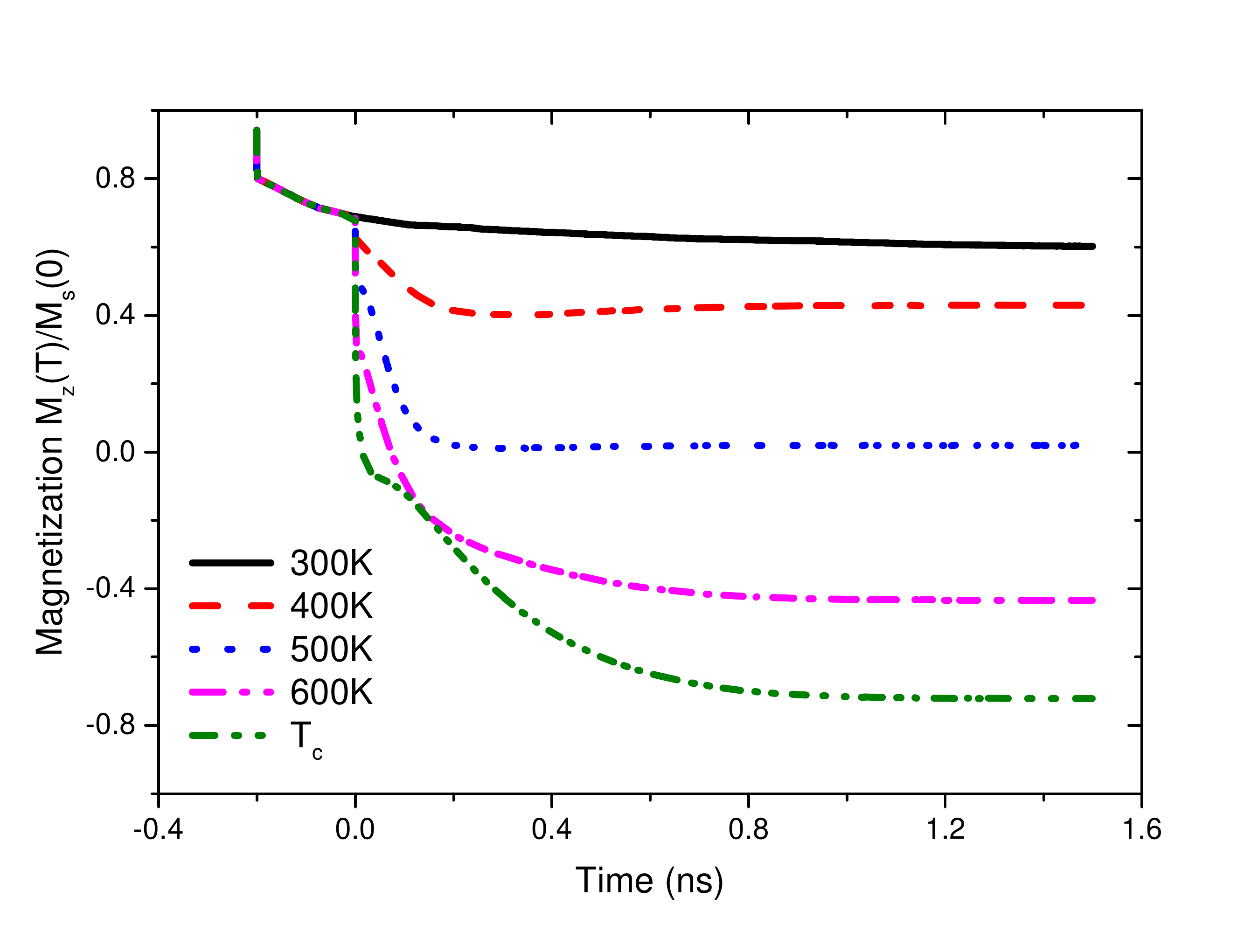}
  \includegraphics[width=8.0cm]{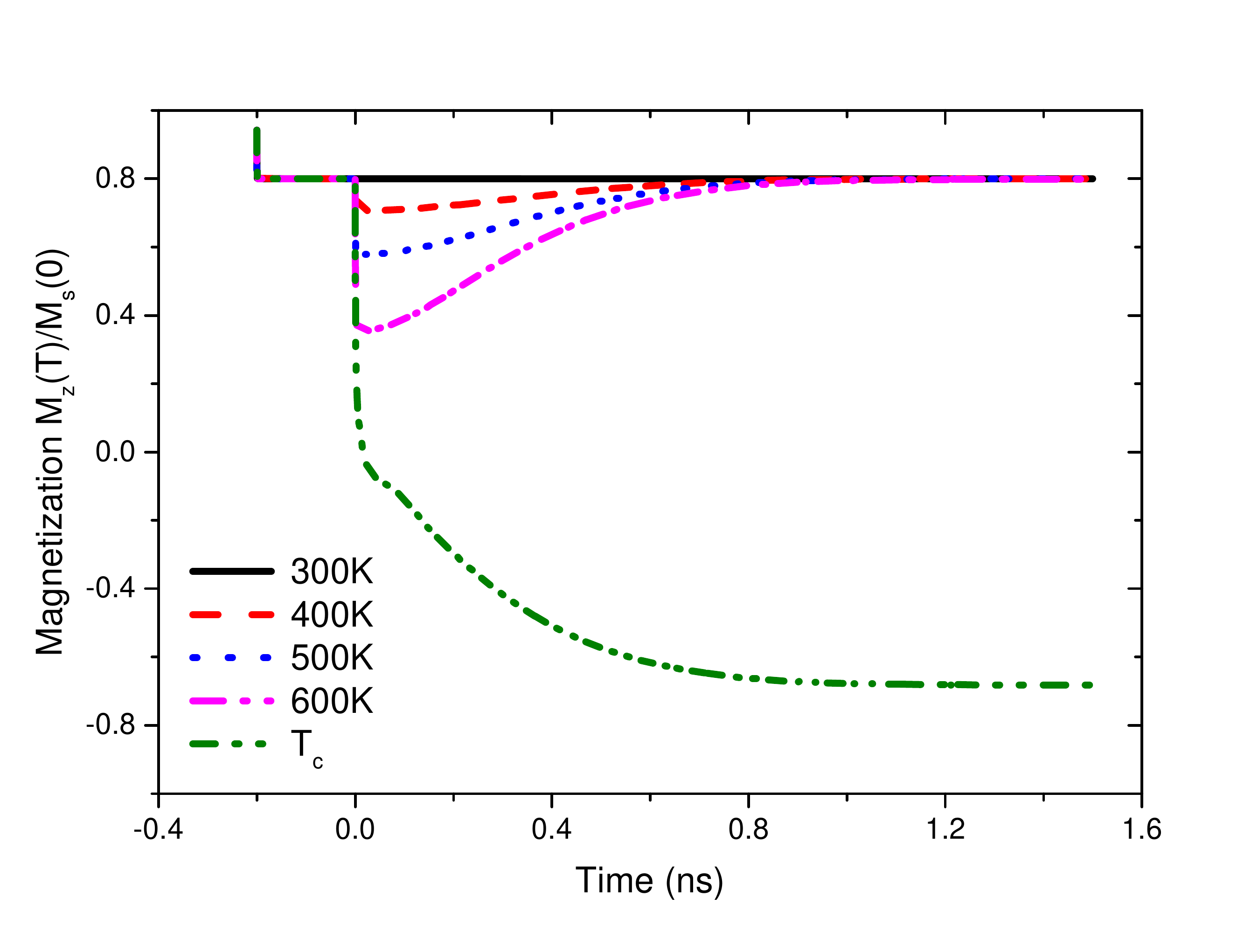}
  \caption{(Color online) Top panel: An M$_{z}$ time sequence for the thin film magnetization, for the 20\% L1$_{o}$ case, with a 6kOe applied field throughout the simulation. The time sequence  shows the effect of applying a temperature increase during the magnetization reversal process and it also shows that the temperature must be increased to $T_{\mathrm{c}}$ to fully reverse the magnetization. Bottom panel: The M$_{z}$ time sequence for the thin film magnetization, for the 100\% L1$_{o}$ case again with a 6kOe applied field. Note the higher anisotropy simulation requires a higher temperature increase to achieve the same level of magnetization reversal.}
  \label{fig:20/100L1otimesequence}
\end{figure}

In order to investigate the underlying switching mechanisms involved in HAMR we use a simplified picture in which the applied field is held constant as the temperature is changed using the same approach as described above when replicating the measurement method of Liu et al\cite{Liu}. The effect of different levels of increased temperature and cooling rate are both determined. Fig.~\ref{fig:20/100L1otimesequence} shows the M$_{z}$ time sequences for the 20\% and 100\% L1$_{o}$ ordered FePt cases, where the applied field strength was 6kOe throughout. For the 20\% FePt case at temperatures well below $T_{\mathrm{c}}$ one can distinguish three distinct time regimes. Initially there is a very rapid reduction of the magnetization, driven by the rapid temperature increase. This involves longitudinal magnetization changes; i.e. the reduction in the magnitude of the magnetization of the grains. The longitudinal relaxation time is of the order of a few hundred fs, as shown by Chubykalo-Fesenko et.al.\cite{Chubykalo}, and as a result the magnetization very rapidly follows the temperature increase. After this process the magnetization proceeds by precessional switching over the thermally reduced energy barriers. Finally, as the temperature lowers there is a slower change in the magnetization caused by thermally activated reversal events which, depending on the grain size, can be significant out to 10-20 ns. The heat assist effect is clear from the data presented. At ambient temperature the field is insufficient to reverse a significant number of grains. As the temperature increases, increasing numbers of grains can be switched due to the reduction in the value of K$_{u}$.

This supports the conventional picture of HAMR as a process driven by thermal activation over thermally reduced energy barriers, and applies to all peak temperatures considered up to 600K. However, it is important to note that during these processes the magnetization is not completely reversed. Complete reversal is observed only for temperatures T $\gtrsim$ 650K. This effect is even more pronounced in calculations for 100\% ordered FePt, as shown in Fig.~\ref{fig:20/100L1otimesequence}, where magnetization reversal is not seen even for temperatures as high as 600K.

\subsection{Linear Reversal}

\begin{figure}[htb]
  \includegraphics[width=8.0cm]{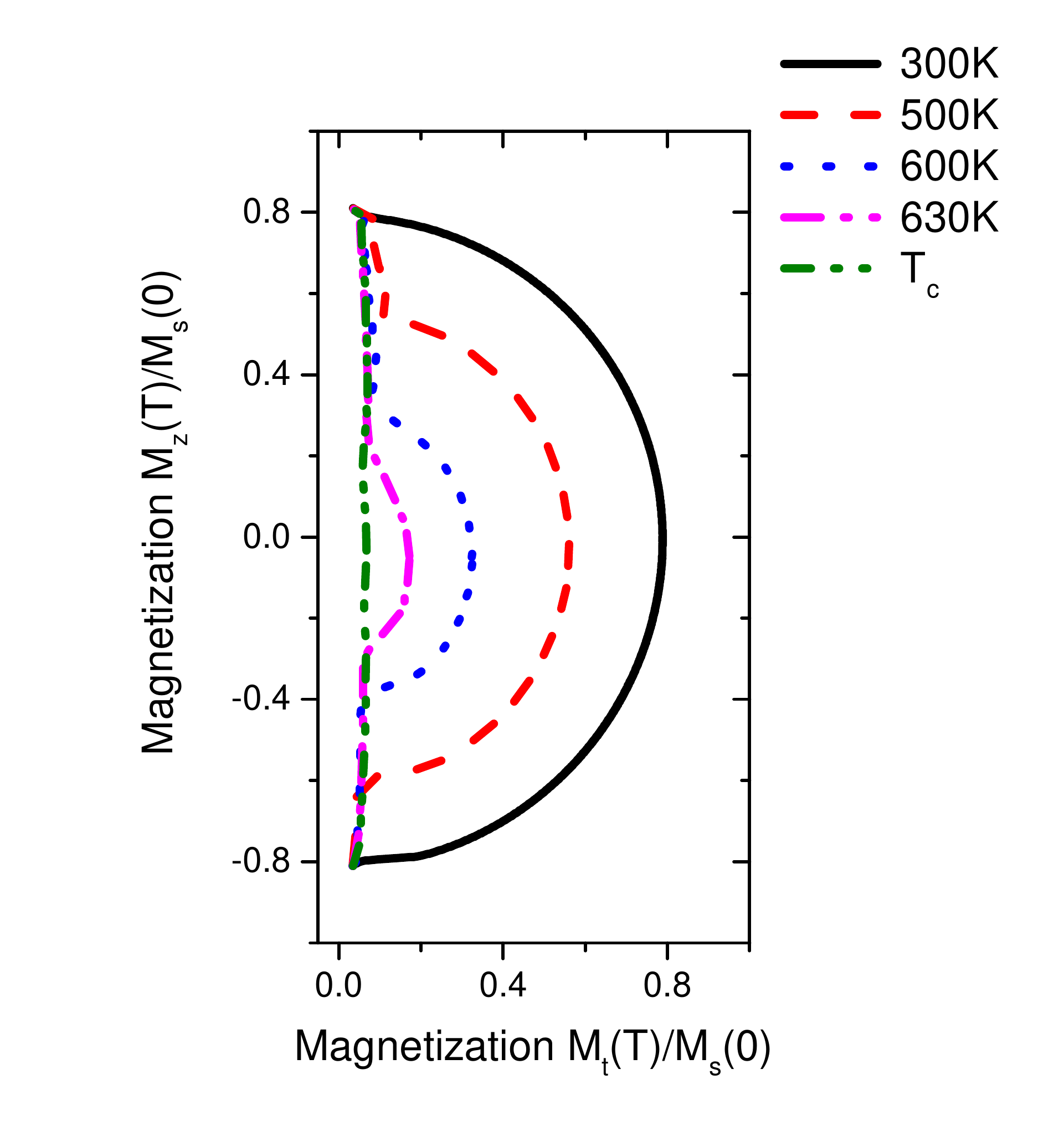}
  \caption{(Color online) The reversal mechanism of an individual grains magnetization (the data is presented as the transverse magnetization component vs the z magnetization component) for the 20\% L1$_{o}$ ordered case, over a range of maximum system temperatures. The change in reversal modes is clearly displayed, from a strictly precessional type with a maximum temperature below 550K, through the elliptical mode as the temperature approaches $T_{\mathrm{c}}$, and onto the linear reversal mode at temperatures above 650K.}
  \label{fig:LinearRev-Hc}
\end{figure}

It is also interesting to note that there is a significant change in behavior between peak temperatures of 640K and 660K. This can be interpreted as arising from the onset of the linear reversal mode\cite{Kazantseva,Barker}; which is characterised by ultrafast reversal via longitudinal magnetization changes. The transition to linear reversal is demonstrated in Fig.~\ref{fig:LinearRev-Hc}, which shows the longitudinal vs transverse magnetization throughout the reversal process for the low anisotropy, 20\% L1$_{o}$ ordered FePt, case. The results are consistent with the atomistic simulations given by Barker et.al.\cite{Barker}. At low temperatures the reversal is by coherent rotation, characterised by a constant magnitude of the magnetization throughout this process, with the magnetization magnitude decreasing with increased temperature. In this regime the reversal is precessional with characteristic switching times in the order of 10-100ps. However, above a temperature of 600K reversal proceeds via a different, non-precessional, switching mechanism termed elliptical and linear reversal. While reversal proceeds by precession of individual spins at the atomistic level, the macroscopic spin associated with individual grains reverses via a reduction of the magnetization and, in the case of linear reversal, collapse of the magnetization through zero.

\begin{figure}[htb]
  \includegraphics[width=8.0cm]{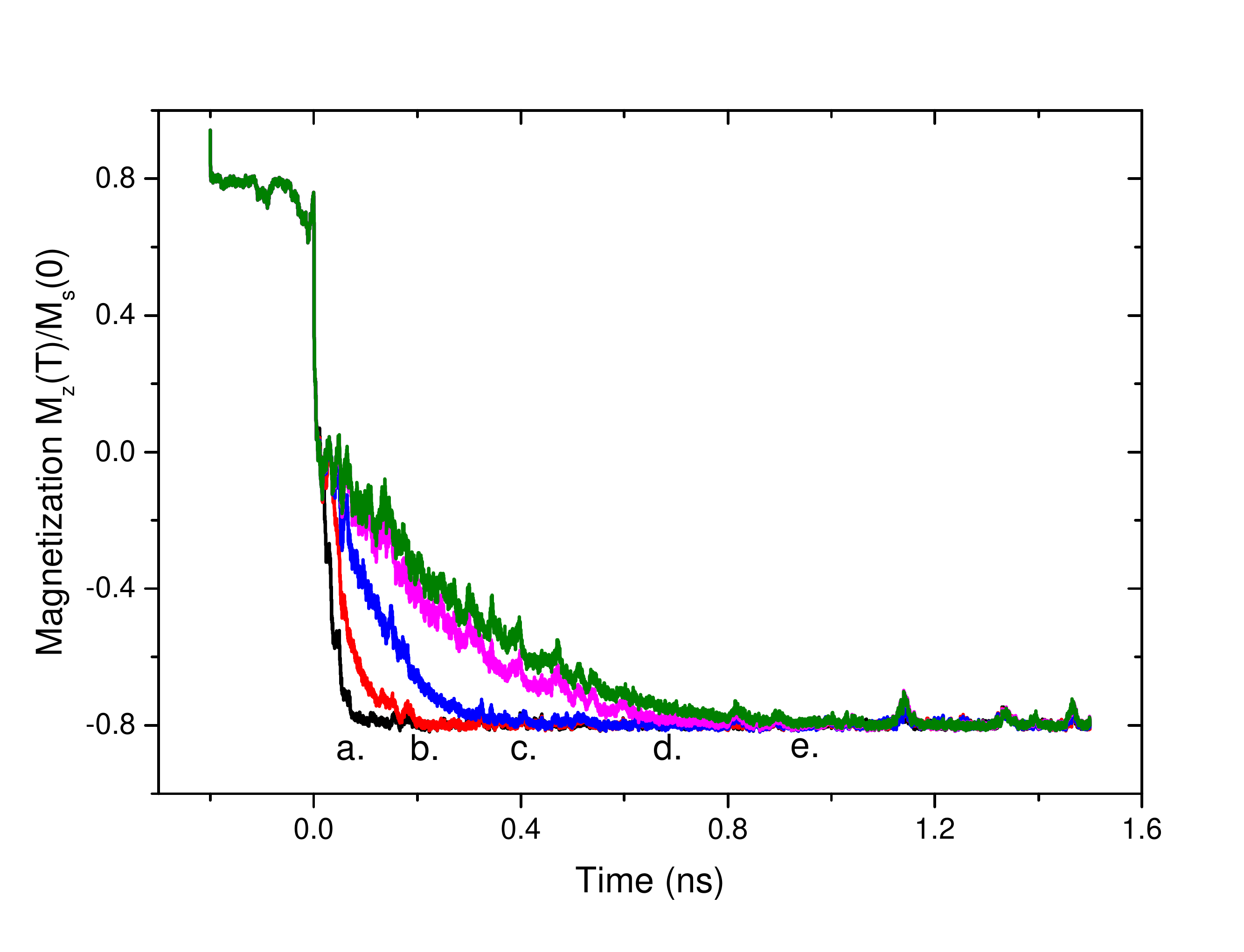}
  \caption{(Color online) The linear reversal case (heating to T$_{c})$, for a single grain, over a range of cooling rates. The reversal of the individual grains magnetization is seen to happen at a greater rate with the quicker cooling time. The cooling times are as follows: a. 0.05ns, b. 0.1ns, c. 0.2ns, d. 0.4ns, e. 0.5ns.}
  \label{fig:TcCoolingrate}
\end{figure}

Within the temperature range of 600K to 640K the magnetization is seen to reverse through the elliptical reversal mode, where the  magnetization decreases to a minimum magnitude in the anisotropy hard direction and then increases as the the system temperature returns to ambient. This process gives the characteristic elliptical reversal pattern seen in Fig.~\ref{fig:LinearRev-Hc}, and importantly results in a reduced energy barrier relative to coherent reversal. Above a temperature of 640K reversal occurs via the linear reversal path; where the magnetization of the individual grains is effectively destroyed on heating and then rebuilds in the direction of the applied field as the system temperature returns to ambient. This mechanism is extremely fast acting with characteristic times of the order of the longitudinal relaxation time of the FePt (100fs).  The dynamical reversal process is illustrated in Fig.~\ref{fig:TcCoolingrate} which shows the M$_{z}$ time sequence for an individual grain, for a peak temperature of $T_{\mathrm{c}}$ and over a range of cooling rates. The magnetization vanishes and reverses rapidly as shown, after which the evolution of the magnetization essentially follows the time evolution of the temperature profile during the cooling process.

\begin{figure}
  \includegraphics[width=8.0cm]{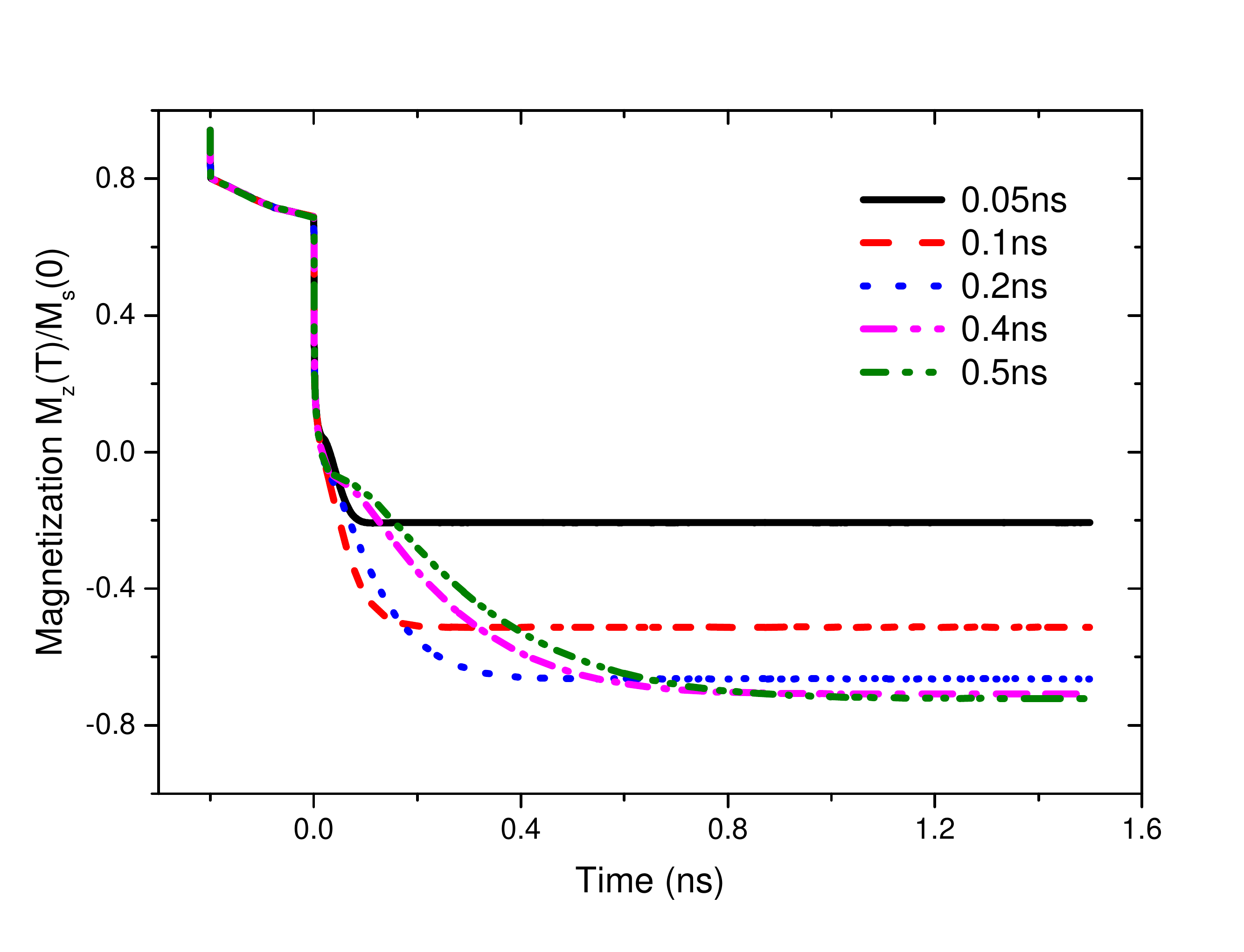}
  \caption{(Color online) The time variation of the magnetization, after  heating to $T_{\mathrm{c}}$, for the thin film magnetization, over a range of cooling rates. For a cooling rate below 0.2ns the level of magnetization reversal is significantly reduced.}
  \label{fig:TccoolingTF}
\end{figure}

We now turn to the behavior of the ensemble of grains, particularly investigating the effect of the cooling rate. The results are given in Fig.~\ref{fig:TccoolingTF}, which shows the time evolution of the magnetization after heating to $T_{\mathrm{c}}$.
For the lowest cooling rate cases, 0.2ns and below, the magnetization of a portion of the grains remains in the initial orientation after the HAMR process. Complete reversal is only achieved for slow cooling rates. We interpret the data as arising from a 'switching window' defined by the time between the appearance of the magnetization on cooling through $T_{\mathrm{c}}$ and the blocking temperature at which the magnetization freezes.

For illustration we first present a simple model of HAMR as a Field Cooled Magnetization (FCM), which is the magnetization achieved after a sample is cooled, from a high temperature state of zero magnetization, through its blocking temperature in an applied field. This is essentially the HAMR recording process, except that the sample may be cooled from above $T_{\mathrm{c}}$. If we assume that the sample attains thermal equilibrium immediately on cooling through $T_{\mathrm{c}}$ we can apply the semi-analytical approach of Chantrell and Wohlfarth~\cite{Chantrell}, with the proviso that at each temperature the saturation magnetization and anisotropy must be re-evaluated. In Ref.~\onlinecite{Chantrell} it was shown that the FCM is dependent on the rate of reduction of the temperature ($\dot{T}$). Essentially, the magnetization is assumed to be in thermal equilibrium above a rate dependent blocking temperature $T_B$, with the equilibrium magnetization being frozen in as the temperature is cooled through $T_B$. In Ref.~\onlinecite{Chantrell} it was shown that the rate dependent blocking temperature is determined via the parameter $Z_c$, given as a solution of the equation:

\begin{equation}
Z_c =\ln ( \dot{T} 25 f_0 T_K (1-h)^2/Z_c^2 )
\label{equ1}
\end{equation}

\noindent where $Z_c = \frac{K_{u}V}{k_{B}T_B}(1-h)^2$, $T_K = K_{u}V/25k_{B}$ is the blocking temperature for quasi-static measurements, and $h$ is the reduced field $H/H_K$. From Eq. \ref{equ1}, $T_B$ is given as:

\begin{equation}
T_B= \frac{K_{u}V}{k_BZ_c}(1-h)^2
\label{tb}
\end{equation}

In Ref.~\onlinecite{Chantrell} the blocking temperature was calculated in an analytical approximation, but here we will use a numerical solution of Eq.~\ref{equ1}. Consider a monodisperse ensemble of particles (diameter $D$) with aligned easy axes, an anisotropy constant K$_{u}$, and saturation magnetization of $M_s$. According to Ref.~\onlinecite{Chantrell}, within the blocking model the system is assumed to be in thermal equilibrium for $T>T_B$, and makes a transition to non-equilibrium behaviour at $T_B$. Given that the thermal equilibrium magnetization for an aligned system is $M=\tanh \beta$ with $\beta = M_s V H/k_{B}T$, we have that:

\begin{equation}
M = \tanh \beta \hspace{1cm}(T>T_B),
\label{abovetb}
\end{equation}

\noindent and

\begin{equation}
M = \tanh \beta_c \hspace{1cm}(T \le T_B),
\label{belowtb}
\end{equation}

\noindent where $\beta_c = M_s V H/k_{B}T_B$. Eq.~\ref{belowtb} reflects the fact that the magnetization below $T_B$ remains at the value frozen in at $T_B$. To calculate $K_{U}(T)$ the Callen-Callen theory  is used. Given the temperature dependences of K$_{u}$ and $M_{s}$ it is straightforward to calculate $M(T)$ using Eqs.~\ref{equ1}-\ref{belowtb}. The FCM has been calculated as a function of the grain size, applied field and cooling rate.

\begin{figure}[htb]
\includegraphics[width=8.cm]{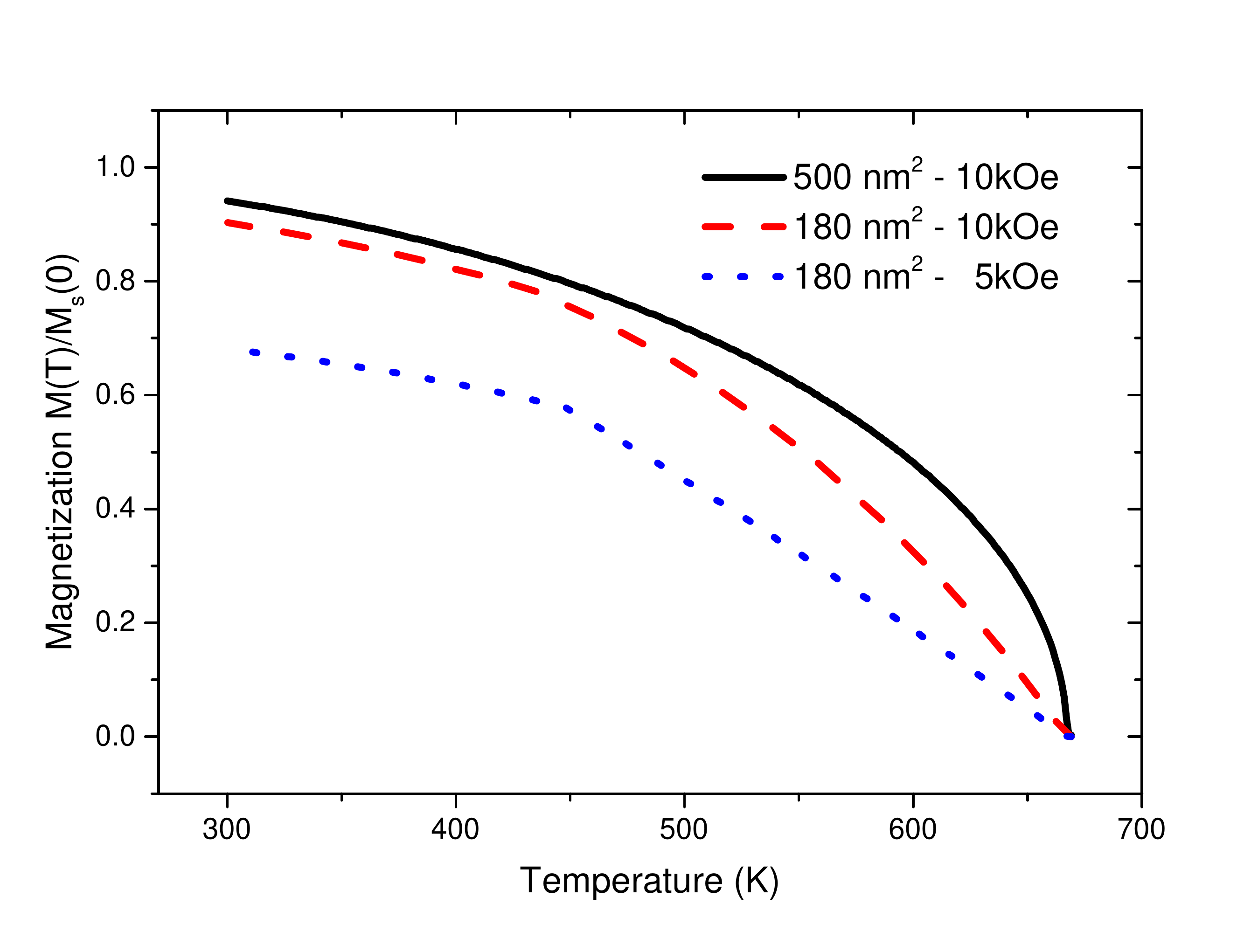}
\caption{(Color online) FCM curves as a function of particle size and magnetic field. The cooling rate is 500Kns$^{-1}$. The system becomes increasingly blocked as the size of both particle and applied field are reduced.}
\label{mT_0vsT_V_H}
\end{figure}

Fig.~\ref{mT_0vsT_V_H} shows the variation of the FCM with temperature for a variety of parameters, and a cooling rate of 500Kns$^{-1}$. The data presented is for a number if particle sizes and applied field strengths. For the 500nm$^{2}$ sized particle in a field of 10kOe the magnetization varies smoothly with temperature. Whereas the smaller particles of 180nm$^{2}$ the magnetization shows a change of slope at around 450K, indicating the onset of blocking. This is more pronounced in the final set of data, in a smaller field of 5kOe. While the semi-analytical approach reproduces the form of the recovery of the magnetization on cooling, it does not reproduce the reduction of the switched magnetization at high sweep rates evident in Fig.~\ref{fig:TccoolingTF}.

This is due to the very rapid cooling under these conditions, which causes the system to go from the high temperature - low anisotropy regime, where magnetization reversal can occur, to the low temperature - high anisotropy regime where reversal is forbidden, in a period of time which is too fast to allow the system to reach thermal equilibrium at each temperature. This result is a dynamic generalization of the concept of 'thermal writability', suggesting that cooling rates should be sufficiently slow as to allow the magnetization to equilibrate. We can quantify this in terms of the freezing time related to the temperature sweep rate $R_T$ as follows; $\tau_f=(T_c-T_B)/R_T$, essentially $\tau_f$ is the time between the Curie temperature and the blocking temperature. For efficient switching in HAMR, $\tau_r>\tau_f$, where $\tau_r$ is the relaxation time of the magnetization during the cooling process. This stresses the importance of the linear reversal mechanism~\cite{Kazantseva_3,Barker_2}, which gives access to relaxation times of the order of picoseconds~\cite{Evans_3}.

\begin{figure}[htb]
  \includegraphics[width=8.0cm]{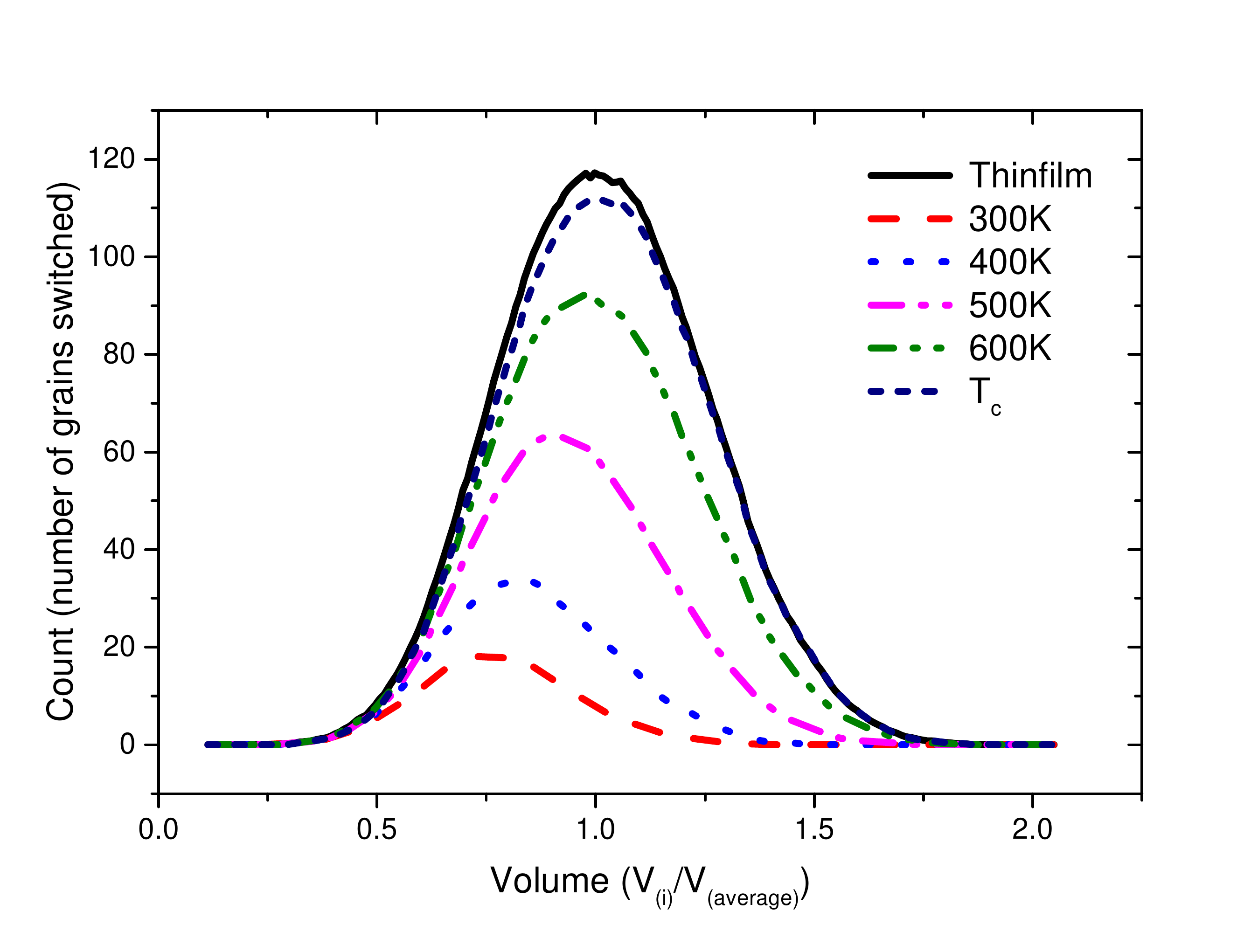}
  \includegraphics[width=8.0cm]{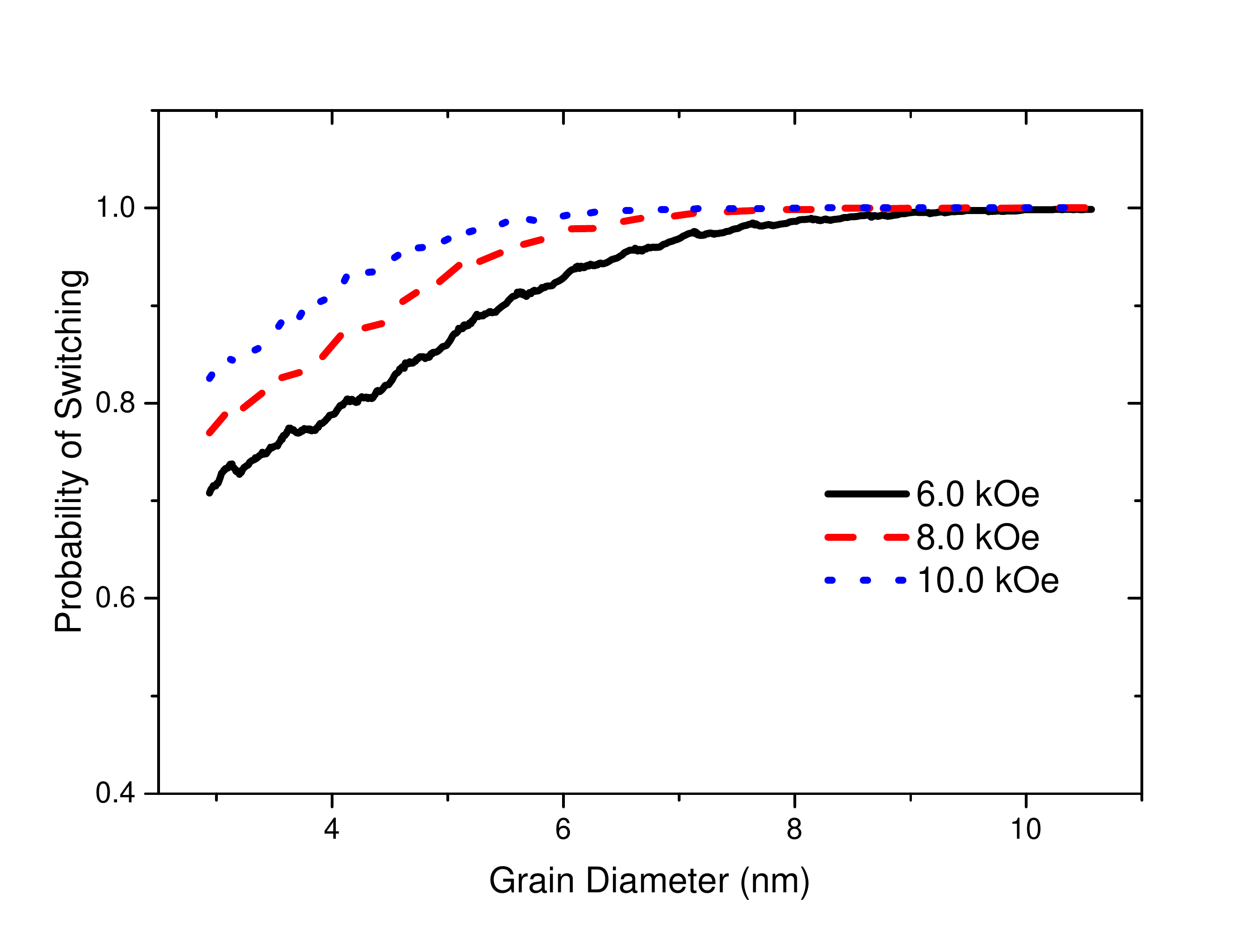}
  \caption{(Color online) Top panel: The spread of the relative volume of the FePt grains within the thin film, showing the spread of the volume for the grains which have reversed the magnetization after the HAMR process, over a range of maximum system temperatures. 'Thin film' denotes the intrinsic size dispersion of the system. Bottom panel: The reversal percentage for the 100\% L1$_{o}$ ordered thin film, where the temperature was raised to $T_{\mathrm{c}}$ during the HAMR process. Below a critical grain size the probability of the magnetization reversing is significantly reduced for a given applied field strength, where the critical grain size is reduced with increased field strength.}
  \label{fig:volumeMzneg}
\end{figure}

\subsection{Grain Volume}

Further insight about the HAMR process is gained by investigating the reversal process as a function of the grain volume and temperature as shown in Fig.~\ref{fig:volumeMzneg}(top panel). Here the number of grains reversed, for a given volume, at each maximum temperature is given. Fig.~\ref{fig:volumeMzneg} demonstrates that the smaller grains are reversed at lower temperatures, with an increasing fraction of the large grains reversed at elevated temperatures. This result is again consistent with the conventional view of HAMR, as increasing the temperature lowers the anisotropy further allowing for ever larger grains to reverse. However it is interesting to note that between peak temperatures of 600K and 660K there is a significant increase in number of grains switch with no corresponding large increase in the mean diameter switched. This is likely due to access to the elliptical and linear reversal modes in this region which will assist the system in reaching thermal equilibrium values for the magnetization.

Our results also demonstrate the effects of the concept of the 'thermal writability'\cite{Evans_2,Richter} which introduces the requirement that the write field H$_{wr}$ should be sufficiently large not only to switch the magnetization, but also to ensure that at the freezing temperature (T$_{f}$) of the grains (M$_{s}$(T)V) is large enough to ensure that there is no thermally-induced back-switching of the magnetization. This is demonstrated in Fig.~\ref{fig:volumeMzneg}(bottom panel) where the probability of reversal is significantly reduced below a critical grain size. The critical grain size decreases with increasing field strength, indicating an increase of the blocking Temperature for fixed grain size. Practically this is notable for two reasons. Firstly it suggests that smaller grains will create increased dc noise. Secondly, we note that the switching field of 10kOe is not sufficient to fully switch grains of less than around 5nm in diameter, which confirms the prediction of Evans et.al.~\cite{Evans_2} that thermal writability is a more stringent criterion than that of thermal stability given that FePt grains as small as around 3.3nm in diameter~\cite{Weller} are thermally stable according to the usual criterion $K_{u}V/(k_{B}T) = 60$.

\subsection{HAMR implications}

\begin{figure}
  \includegraphics[width=8.0cm]{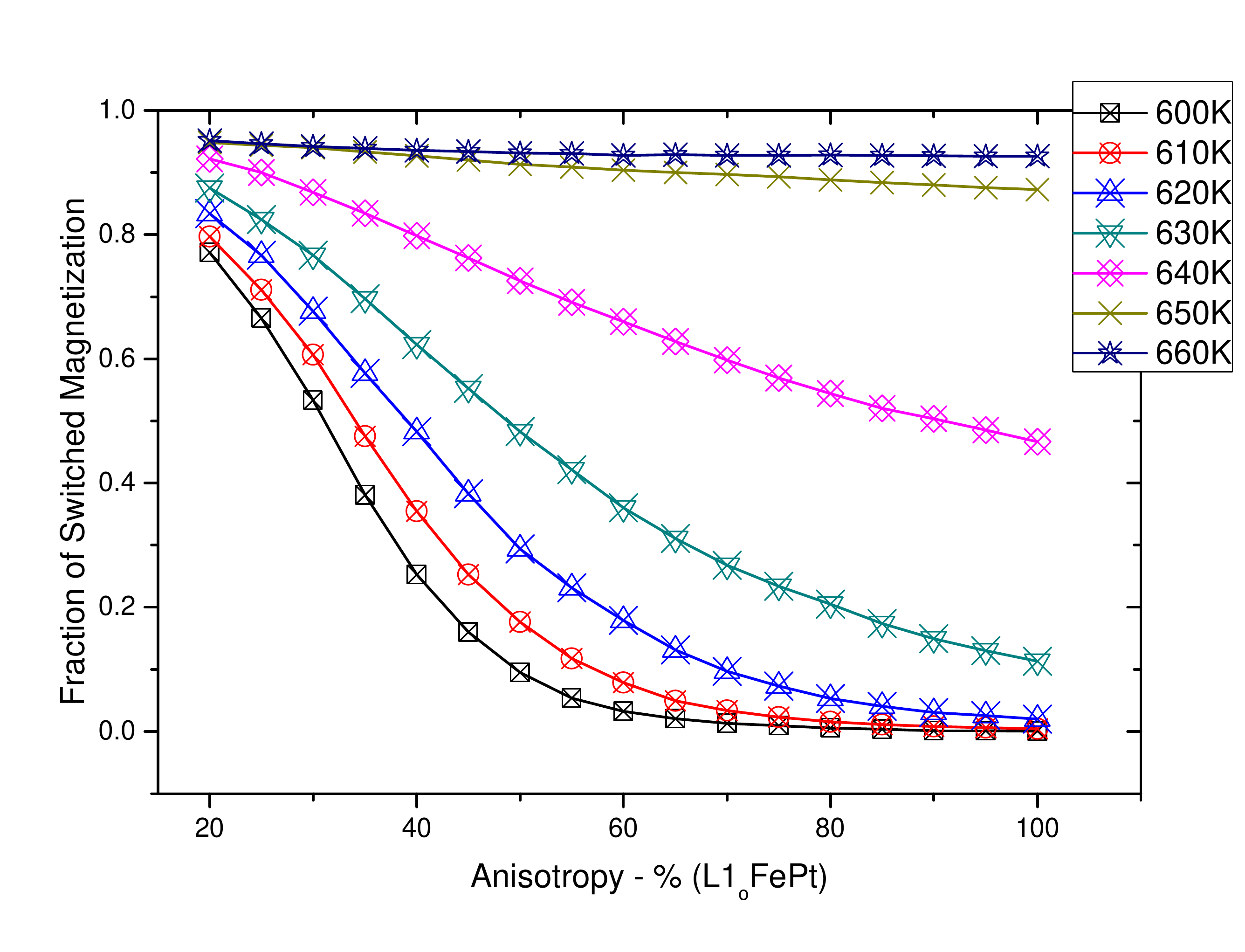}
  \caption{(Color online) The reversal probability of the thin film magnetization with a 6kOe applied field, over a range of maximum system temperatures (600K - $T_{\mathrm{c}}$) and anisotropies (20\% - 100\% Bulk L1$_{o}$ ordered FePt grains).}
  \label{fig:ReversalProbability}
\end{figure}

In terms of applications in recording, the final state of the system after the HAMR process is of central importance. Here we characterise the final state in terms of a reversal probability, taken as the reduced magnetization (relative to saturation), after the HAMR process. Fig.~\ref{fig:ReversalProbability} shows the reversal probability over a range of anisotropy values and maximum system temperatures. This clearly shows the increase in the temperature required to initiate the reversal process in the higher anisotropy simulations. Above an anisotropy of 60\% L1$_{o}$ ordered grains and below a temperature of 600K the system shows very low or zero magnetization reversal. The data also show that, when the maximum system temperature reaches $T_{\mathrm{c}}$, reversal is achieved for all anisotropy values. For the relatively low anisotropy values it can be seen that there is a relatively weak dependence of the reversal probability, although the probability does increase up to $T_{\mathrm{c}}$. There is a much more dramatic effect of temperature for values of anisotropy approaching that of Bulk FePt. Here, peak temperatures as high as 640K are not sufficient to cause switching of the magnetization; a switching path is only available via linear reversal, which suggests that this mechanism will increasingly become the dominant process as the anisotropy value increases in line with the requirement of thermal stability at ultra-high densities.

These two constraints; the higher anisotropy regimes needing increasingly higher temperatures (approaching $T_{\mathrm{c}}$) to initiate reversal, means that the temperature range where reversal is seen is reduced as the anisotropy is raised. This is a significant result for HAMR, as a narrower range of temperature where magnetization reversal is seen implies the reversal process must take place over a smaller timescale dictated by the timescale where the system temperature is at $T_{\mathrm{c}}$. This could lead to the possibility of incomplete switching, as reversal must occur at this high temperature. In this respect, the intrinsically fast (ps timescale) of switching via linear reversal is an important factor.

\section{Conclusion}

We have investigated the physics the HAMR process; a viable extension to hard disk magnetic data storage technologies. As expected the field required to reverse the magnetization of a given granular thin film sample is reduced with increased temperature. However, we have shown that the HAMR process is not simply magnetization reversal over a thermally reduced energy barrier, in the high temperature/low anisotropy regime, and is far more complex. To reverse the magnetization via the HAMR process requires that the temperature be raised close to $T_{\mathrm{c}}$ for all anisotropies investigated and that the write field ($H_{wr}$) must be sufficient not only to reverse the magnetization, but also to overcome backswitching due to the excess thermal energy associated with the HAMR process.

At temperatures sufficiently below $T_{\mathrm{c}}$, in the case of FePt T$\le 600$K, the individual grains are seen to reverse through 3 stages; firstly a reduction in the magnitude of the magnetization associated with the rapid temperature rise. Then reversal occurs over the reduced anisotropy, as in the conventional view of HAMR, where the anisotropy has been sufficiently reduced to allow precessional reversal over the reduced energy barrier. Finally, as the temperature lowers there is a slower change in the magnetization caused by thermally activated reversal events which, depending on the grain size, can be significant out to 10-20ns. These effect are anisotropy dependent and therefore the temperatures where these effects are seen are much larger for the higher anisotropy regimes and in this fashion the magnetization of the thin film is not fully reversed. At higher temperatures the linear reversal comes into play, giving a faster route to achieve the thermal equilibrium magnetization, which thermodynamically is the maximum possible~\cite{Evans_2}.

We have demonstrated that to set the magnetization to the preferred direction, via the HAMR process,  the temperature must be raised to $T_{\mathrm{c}}$ and the applied field must be sufficient to overcome the superparamagnetic freezing associated with the rapid temperature decrease. This implies that magnetic recording, via HAMR, is only possible through the newly proposed linear reversal mode, where the magnetization is entirely destroyed by the temperature increase to $T_{\mathrm{c}}$ and then is rebuilt in the direction of the applied field as the temperature returns to ambient. We find that the thermal writability factor~\cite{Evans_2} is an important limitation in HAMR. Our simulations show that at diameters below around 5nm backswitching becomes increasingly important, suggesting that thermal writability is a more stringent limitation than thermal stability since fully ordered grains as small as 3.3nm are expected to be thermally stable. However, we note that our calculations assume a  value of 10kOe for the switching field. While this is realistic for current technology, it suggests that the introduction of higher effective fields such as are available through, for example, thermally induced magnetization switching\cite{Ostler} will be necessary for magnetic recording to achieve its full potential.

\section*{Acknowledgements}

Financial support of the Advanced Storage Technology Consortium is gratefully acknowledged. This work made use of the facilities of N8 HPC provided and funded by the N8 consortium and EPSRC (Grant No. EP/K000225/1) co-ordinated by the Universities of Leeds and Manchester and the EPSRC Small items of research equipment at the University of York ENERGY (Grant No. EP/K031589/1). The advice of GP Ju, KK Wang and PW Huang is greatfully acknowledged.


\begin{thebibliography}{23}%
\makeatletter
\providecommand \@ifxundefined [1]{%
 \@ifx{#1\undefined}
}%
\providecommand \@ifnum [1]{%
 \ifnum #1\expandafter \@firstoftwo
 \else \expandafter \@secondoftwo
 \fi
}%
\providecommand \@ifx [1]{%
 \ifx #1\expandafter \@firstoftwo
 \else \expandafter \@secondoftwo
 \fi
}%
\providecommand \natexlab [1]{#1}%
\providecommand \enquote  [1]{``#1''}%
\providecommand \bibnamefont  [1]{#1}%
\providecommand \bibfnamefont [1]{#1}%
\providecommand \citenamefont [1]{#1}%
\providecommand \href@noop [0]{\@secondoftwo}%
\providecommand \href [0]{\begingroup \@sanitize@url \@href}%
\providecommand \@href[1]{\@@startlink{#1}\@@href}%
\providecommand \@@href[1]{\endgroup#1\@@endlink}%
\providecommand \@sanitize@url [0]{\catcode `\\12\catcode `\$12\catcode
  `\&12\catcode `\#12\catcode `\^12\catcode `\_12\catcode `\%12\relax}%
\providecommand \@@startlink[1]{}%
\providecommand \@@endlink[0]{}%
\providecommand \url  [0]{\begingroup\@sanitize@url \@url }%
\providecommand \@url [1]{\endgroup\@href {#1}{\urlprefix }}%
\providecommand \urlprefix  [0]{URL }%
\providecommand \Eprint [0]{\href }%
\providecommand \doibase [0]{http://dx.doi.org/}%
\providecommand \selectlanguage [0]{\@gobble}%
\providecommand \bibinfo  [0]{\@secondoftwo}%
\providecommand \bibfield  [0]{\@secondoftwo}%
\providecommand \translation [1]{[#1]}%
\providecommand \BibitemOpen [0]{}%
\providecommand \bibitemStop [0]{}%
\providecommand \bibitemNoStop [0]{.\EOS\space}%
\providecommand \EOS [0]{\spacefactor3000\relax}%
\providecommand \BibitemShut  [1]{\csname bibitem#1\endcsname}%
\let\auto@bib@innerbib\@empty
\bibitem [{\citenamefont {McDaniel}(2005)}]{McDaniel}%
  \BibitemOpen
  \bibfield  {author} {\bibinfo {author} {\bibfnamefont {T.~W.}\ \bibnamefont
  {McDaniel}},\ }\href@noop {} {\bibfield  {journal} {\bibinfo  {journal} {J.
  Phys. Condens. Matter.}\ }\textbf {\bibinfo {volume} {17}},\ \bibinfo {pages}
  {R315} (\bibinfo {year} {2005})}\BibitemShut {NoStop}%
\bibitem [{\citenamefont {Wood}(2000)}]{Wood}%
  \BibitemOpen
  \bibfield  {author} {\bibinfo {author} {\bibfnamefont {R.}~\bibnamefont
  {Wood}},\ }\href@noop {} {\bibfield  {journal} {\bibinfo  {journal} {IEEE
  Trans. Mag.}\ }\textbf {\bibinfo {volume} {36}},\ \bibinfo {pages} {36}
  (\bibinfo {year} {2000})}\BibitemShut {NoStop}%
\bibitem [{\citenamefont {Bean}(1959)}]{Bean}%
  \BibitemOpen
  \bibfield  {author} {\bibinfo {author} {\bibfnamefont {J.}~\bibnamefont
  {Bean}, \bibfnamefont {C.~P.~Livingston}},\ }\href@noop {} {\bibfield
  {journal} {\bibinfo  {journal} {J. Apl. Phys}\ }\textbf {\bibinfo {volume}
  {S30}},\ \bibinfo {pages} {120S} (\bibinfo {year} {1959})}\BibitemShut
  {NoStop}%
\bibitem [{\citenamefont {Granz}\ and\ \citenamefont {Kryder}(2012)}]{Granz}%
  \BibitemOpen
  \bibfield  {author} {\bibinfo {author} {\bibfnamefont {S.~D.}\ \bibnamefont
  {Granz}}\ and\ \bibinfo {author} {\bibfnamefont {M.~H.}\ \bibnamefont
  {Kryder}},\ }\href@noop {} {\bibfield  {journal} {\bibinfo  {journal} {J.
  Magn. Magn. Matter.}\ }\textbf {\bibinfo {volume} {324}},\ \bibinfo {pages}
  {287 } (\bibinfo {year} {2012})}\BibitemShut {NoStop}%
\bibitem [{\citenamefont {Heinonen}\ and\ \citenamefont
  {Gao}(2008)}]{Heinonen}%
  \BibitemOpen
  \bibfield  {author} {\bibinfo {author} {\bibfnamefont {O.}~\bibnamefont
  {Heinonen}}\ and\ \bibinfo {author} {\bibfnamefont {K.~Z.}\ \bibnamefont
  {Gao}},\ }\href@noop {} {\bibfield  {journal} {\bibinfo  {journal} {J. Magn.
  Magn. Matter}\ }\textbf {\bibinfo {volume} {320}},\ \bibinfo {pages} {2885}
  (\bibinfo {year} {2008})}\BibitemShut {NoStop}%
\bibitem [{\citenamefont {Ju}\ \emph {et~al.}(2015)\citenamefont {Ju},
  \citenamefont {Peng}, \citenamefont {Chang}, \citenamefont {Ding},
  \citenamefont {Wu}, \citenamefont {Zhu}, \citenamefont {Kubota},
  \citenamefont {Klemmer}, \citenamefont {Amini}, \citenamefont {Gao},
  \citenamefont {Fan}, \citenamefont {Rausch}, \citenamefont {Subedi},
  \citenamefont {Ma}, \citenamefont {Kalarickal}, \citenamefont {Rea},
  \citenamefont {Dimitrov}, \citenamefont {Huang}, \citenamefont {Wang},
  \citenamefont {Chen}, \citenamefont {Peng}, \citenamefont {Chen},
  \citenamefont {Dykes}, \citenamefont {Seigler}, \citenamefont {Gage},
  \citenamefont {Chantrell},\ and\ \citenamefont {Thiele}}]{Ju}%
  \BibitemOpen
  \bibfield  {author} {\bibinfo {author} {\bibfnamefont {G.}~\bibnamefont
  {Ju}}, \bibinfo {author} {\bibfnamefont {Y.}~\bibnamefont {Peng}}, \bibinfo
  {author} {\bibfnamefont {E.~K.~C.}\ \bibnamefont {Chang}}, \bibinfo {author}
  {\bibfnamefont {Y.}~\bibnamefont {Ding}}, \bibinfo {author} {\bibfnamefont
  {A.~Q.}\ \bibnamefont {Wu}}, \bibinfo {author} {\bibfnamefont
  {X.}~\bibnamefont {Zhu}}, \bibinfo {author} {\bibfnamefont {Y.}~\bibnamefont
  {Kubota}}, \bibinfo {author} {\bibfnamefont {T.~J.}\ \bibnamefont {Klemmer}},
  \bibinfo {author} {\bibfnamefont {H.}~\bibnamefont {Amini}}, \bibinfo
  {author} {\bibfnamefont {L.}~\bibnamefont {Gao}}, \bibinfo {author}
  {\bibfnamefont {Z.}~\bibnamefont {Fan}}, \bibinfo {author} {\bibfnamefont
  {T.}~\bibnamefont {Rausch}}, \bibinfo {author} {\bibfnamefont
  {P.}~\bibnamefont {Subedi}}, \bibinfo {author} {\bibfnamefont
  {M.}~\bibnamefont {Ma}}, \bibinfo {author} {\bibfnamefont {S.}~\bibnamefont
  {Kalarickal}}, \bibinfo {author} {\bibfnamefont {C.~J.}\ \bibnamefont {Rea}},
  \bibinfo {author} {\bibfnamefont {D.~V.}\ \bibnamefont {Dimitrov}}, \bibinfo
  {author} {\bibfnamefont {P.~W.}\ \bibnamefont {Huang}}, \bibinfo {author}
  {\bibfnamefont {K.}~\bibnamefont {Wang}}, \bibinfo {author} {\bibfnamefont
  {X.}~\bibnamefont {Chen}}, \bibinfo {author} {\bibfnamefont {C.}~\bibnamefont
  {Peng}}, \bibinfo {author} {\bibfnamefont {W.}~\bibnamefont {Chen}}, \bibinfo
  {author} {\bibfnamefont {J.~W.}\ \bibnamefont {Dykes}}, \bibinfo {author}
  {\bibfnamefont {M.~A.}\ \bibnamefont {Seigler}}, \bibinfo {author}
  {\bibfnamefont {E.~C.}\ \bibnamefont {Gage}}, \bibinfo {author}
  {\bibfnamefont {R.}~\bibnamefont {Chantrell}}, \ and\ \bibinfo {author}
  {\bibfnamefont {J.~U.}\ \bibnamefont {Thiele}},\ }\href@noop {} {\bibfield
  {journal} {\bibinfo  {journal} {IEEE Trans. Mag.}\ }\textbf {\bibinfo
  {volume} {51}},\ \bibinfo {pages} {1} (\bibinfo {year} {2015})}\BibitemShut
  {NoStop}%
\bibitem [{\citenamefont {Chubykalo-Fesenko}\ \emph {et~al.}(2006)\citenamefont
  {Chubykalo-Fesenko}, \citenamefont {Nowak}, \citenamefont {Chantrell},\ and\
  \citenamefont {Garanin}}]{Chubykalo}%
  \BibitemOpen
  \bibfield  {author} {\bibinfo {author} {\bibfnamefont {O.}~\bibnamefont
  {Chubykalo-Fesenko}}, \bibinfo {author} {\bibfnamefont {U.}~\bibnamefont
  {Nowak}}, \bibinfo {author} {\bibfnamefont {R.~W.}\ \bibnamefont
  {Chantrell}}, \ and\ \bibinfo {author} {\bibfnamefont {D.}~\bibnamefont
  {Garanin}},\ }\href@noop {} {\bibfield  {journal} {\bibinfo  {journal} {Phys.
  Rev. B}\ }\textbf {\bibinfo {volume} {74}},\ \bibinfo {pages} {094436}
  (\bibinfo {year} {2006})}\BibitemShut {NoStop}%
\bibitem [{\citenamefont {Kazantseva}\ \emph
  {et~al.}(2009{\natexlab{a}})\citenamefont {Kazantseva}, \citenamefont
  {Hinzke}, \citenamefont {Chantrell},\ and\ \citenamefont
  {Nowak}}]{Kazantseva}%
  \BibitemOpen
  \bibfield  {author} {\bibinfo {author} {\bibfnamefont {N.}~\bibnamefont
  {Kazantseva}}, \bibinfo {author} {\bibfnamefont {D.}~\bibnamefont {Hinzke}},
  \bibinfo {author} {\bibfnamefont {R.~W.}\ \bibnamefont {Chantrell}}, \ and\
  \bibinfo {author} {\bibfnamefont {U.}~\bibnamefont {Nowak}},\ }\href@noop {}
  {\bibfield  {journal} {\bibinfo  {journal} {EPL (Euro. Phys. Lett.)}\
  }\textbf {\bibinfo {volume} {86}},\ \bibinfo {pages} {27006} (\bibinfo {year}
  {2009}{\natexlab{a}})}\BibitemShut {NoStop}%
\bibitem [{\citenamefont {Barker}\ \emph
  {et~al.}(2010{\natexlab{a}})\citenamefont {Barker}, \citenamefont {Evans},
  \citenamefont {Chantrell}, \citenamefont {Hinzke},\ and\ \citenamefont
  {Nowak}}]{Barker}%
  \BibitemOpen
  \bibfield  {author} {\bibinfo {author} {\bibfnamefont {J.}~\bibnamefont
  {Barker}}, \bibinfo {author} {\bibfnamefont {R.~F.~L.}\ \bibnamefont
  {Evans}}, \bibinfo {author} {\bibfnamefont {R.~W.}\ \bibnamefont
  {Chantrell}}, \bibinfo {author} {\bibfnamefont {D.}~\bibnamefont {Hinzke}}, \
  and\ \bibinfo {author} {\bibfnamefont {U.}~\bibnamefont {Nowak}},\
  }\href@noop {} {\bibfield  {journal} {\bibinfo  {journal} {Appl. Phys.
  Lett.}\ }\textbf {\bibinfo {volume} {97}},\ \bibinfo {eid} {192504} (\bibinfo
  {year} {2010}{\natexlab{a}})}\BibitemShut {NoStop}%
\bibitem [{\citenamefont {Garanin}(1997)}]{Garanin}%
  \BibitemOpen
  \bibfield  {author} {\bibinfo {author} {\bibfnamefont {D.~A.}\ \bibnamefont
  {Garanin}},\ }\href@noop {} {\bibfield  {journal} {\bibinfo  {journal} {Phys.
  Rev. B}\ }\textbf {\bibinfo {volume} {55}},\ \bibinfo {pages} {3050}
  (\bibinfo {year} {1997})}\BibitemShut {NoStop}%
\bibitem [{\citenamefont {Garanin}\ and\ \citenamefont
  {Chubykalo-Fesenko}(2004)}]{Garanin_2}%
  \BibitemOpen
  \bibfield  {author} {\bibinfo {author} {\bibfnamefont {D.~A.}\ \bibnamefont
  {Garanin}}\ and\ \bibinfo {author} {\bibfnamefont {O.}~\bibnamefont
  {Chubykalo-Fesenko}},\ }\href@noop {} {\bibfield  {journal} {\bibinfo
  {journal} {Phys. Rev. B}\ }\textbf {\bibinfo {volume} {70}},\ \bibinfo
  {pages} {212409} (\bibinfo {year} {2004})}\BibitemShut {NoStop}%
\bibitem [{\citenamefont {Evans}\ \emph
  {et~al.}(2012{\natexlab{a}})\citenamefont {Evans}, \citenamefont {Hinzke},
  \citenamefont {Atxitia}, \citenamefont {Nowak}, \citenamefont {Chantrell},\
  and\ \citenamefont {Chubykalo-Fesenko}}]{Evans}%
  \BibitemOpen
  \bibfield  {author} {\bibinfo {author} {\bibfnamefont {R.~F.~L.}\
  \bibnamefont {Evans}}, \bibinfo {author} {\bibfnamefont {D.}~\bibnamefont
  {Hinzke}}, \bibinfo {author} {\bibfnamefont {U.}~\bibnamefont {Atxitia}},
  \bibinfo {author} {\bibfnamefont {U.}~\bibnamefont {Nowak}}, \bibinfo
  {author} {\bibfnamefont {R.~W.}\ \bibnamefont {Chantrell}}, \ and\ \bibinfo
  {author} {\bibfnamefont {O.}~\bibnamefont {Chubykalo-Fesenko}},\ }\href@noop
  {} {\bibfield  {journal} {\bibinfo  {journal} {Phys. Rev. B}\ }\textbf
  {\bibinfo {volume} {85}},\ \bibinfo {pages} {014433} (\bibinfo {year}
  {2012}{\natexlab{a}})}\BibitemShut {NoStop}%
\bibitem [{\citenamefont {Kazantseva}\ \emph {et~al.}(2008)\citenamefont
  {Kazantseva}, \citenamefont {Hinzke}, \citenamefont {Nowak}, \citenamefont
  {Chantrell}, \citenamefont {Atxitia},\ and\ \citenamefont
  {Chubykalo-Fesenko}}]{Kazantseva_2}%
  \BibitemOpen
  \bibfield  {author} {\bibinfo {author} {\bibfnamefont {N.}~\bibnamefont
  {Kazantseva}}, \bibinfo {author} {\bibfnamefont {D.}~\bibnamefont {Hinzke}},
  \bibinfo {author} {\bibfnamefont {U.}~\bibnamefont {Nowak}}, \bibinfo
  {author} {\bibfnamefont {R.~W.}\ \bibnamefont {Chantrell}}, \bibinfo {author}
  {\bibfnamefont {U.}~\bibnamefont {Atxitia}}, \ and\ \bibinfo {author}
  {\bibfnamefont {O.}~\bibnamefont {Chubykalo-Fesenko}},\ }\href@noop {}
  {\bibfield  {journal} {\bibinfo  {journal} {Phys. Rev. B}\ }\textbf {\bibinfo
  {volume} {77}},\ \bibinfo {pages} {184428} (\bibinfo {year}
  {2008})}\BibitemShut {NoStop}%
\bibitem [{\citenamefont {Peng}\ \emph {et~al.}(2011)\citenamefont {Peng},
  \citenamefont {Wu}, \citenamefont {Pressesky}, \citenamefont {Ju},
  \citenamefont {Scholz},\ and\ \citenamefont {Chantrell}}]{Peng}%
  \BibitemOpen
  \bibfield  {author} {\bibinfo {author} {\bibfnamefont {Y.}~\bibnamefont
  {Peng}}, \bibinfo {author} {\bibfnamefont {X.~W.}\ \bibnamefont {Wu}},
  \bibinfo {author} {\bibfnamefont {J.}~\bibnamefont {Pressesky}}, \bibinfo
  {author} {\bibfnamefont {G.~P.}\ \bibnamefont {Ju}}, \bibinfo {author}
  {\bibfnamefont {W.}~\bibnamefont {Scholz}}, \ and\ \bibinfo {author}
  {\bibfnamefont {R.~W.}\ \bibnamefont {Chantrell}},\ }\href@noop {} {\bibfield
   {journal} {\bibinfo  {journal} {J. App. Phys.}\ }\textbf {\bibinfo {volume}
  {109}},\ \bibinfo {eid} {123907} (\bibinfo {year} {2011})}\BibitemShut
  {NoStop}%
\bibitem [{\citenamefont {Evans}\ \emph
  {et~al.}(2012{\natexlab{b}})\citenamefont {Evans}, \citenamefont {Chantrell},
  \citenamefont {Nowak}, \citenamefont {Lyberatos},\ and\ \citenamefont
  {Richter}}]{Evans_2}%
  \BibitemOpen
  \bibfield  {author} {\bibinfo {author} {\bibfnamefont {R.~F.~L.}\
  \bibnamefont {Evans}}, \bibinfo {author} {\bibfnamefont {R.~W.}\ \bibnamefont
  {Chantrell}}, \bibinfo {author} {\bibfnamefont {U.}~\bibnamefont {Nowak}},
  \bibinfo {author} {\bibfnamefont {A.}~\bibnamefont {Lyberatos}}, \ and\
  \bibinfo {author} {\bibfnamefont {H.}~\bibnamefont {Richter}},\ }\href@noop
  {} {\bibfield  {journal} {\bibinfo  {journal} {Appl. Phys. Lett.}\ }\textbf
  {\bibinfo {volume} {100}},\ \bibinfo {eid} {102402} (\bibinfo {year}
  {2012}{\natexlab{b}})}\BibitemShut {NoStop}%
\bibitem [{\citenamefont {Richter}\ \emph {et~al.}(2012)\citenamefont
  {Richter}, \citenamefont {Lyberatos}, \citenamefont {Nowak}, \citenamefont
  {Evans},\ and\ \citenamefont {Chantrell}}]{Richter}%
  \BibitemOpen
  \bibfield  {author} {\bibinfo {author} {\bibfnamefont {H.~J.}\ \bibnamefont
  {Richter}}, \bibinfo {author} {\bibfnamefont {A.}~\bibnamefont {Lyberatos}},
  \bibinfo {author} {\bibfnamefont {U.}~\bibnamefont {Nowak}}, \bibinfo
  {author} {\bibfnamefont {R.~F.~L.}\ \bibnamefont {Evans}}, \ and\ \bibinfo
  {author} {\bibfnamefont {R.~W.}\ \bibnamefont {Chantrell}},\ }\href@noop {}
  {\bibfield  {journal} {\bibinfo  {journal} {J. Appl. Phys.}\ }\textbf
  {\bibinfo {volume} {111}},\ \bibinfo {eid} {033909} (\bibinfo {year}
  {2012})}\BibitemShut {NoStop}%
\bibitem [{\citenamefont {Liu}\ \emph {et~al.}(2009)\citenamefont {Liu},
  \citenamefont {Xu}, \citenamefont {Gao}, \citenamefont {Chen}, \citenamefont
  {Lai}, \citenamefont {Du},\ and\ \citenamefont {Zhou}}]{Liu}%
  \BibitemOpen
  \bibfield  {author} {\bibinfo {author} {\bibfnamefont {X.~D.}\ \bibnamefont
  {Liu}}, \bibinfo {author} {\bibfnamefont {Z.}~\bibnamefont {Xu}}, \bibinfo
  {author} {\bibfnamefont {R.~X.}\ \bibnamefont {Gao}}, \bibinfo {author}
  {\bibfnamefont {Z.~F.}\ \bibnamefont {Chen}}, \bibinfo {author}
  {\bibfnamefont {T.~S.}\ \bibnamefont {Lai}}, \bibinfo {author} {\bibfnamefont
  {J.}~\bibnamefont {Du}}, \ and\ \bibinfo {author} {\bibfnamefont {S.~M.}\
  \bibnamefont {Zhou}},\ }\href@noop {} {\bibfield  {journal} {\bibinfo
  {journal} {J. Appl. Phys.}\ }\textbf {\bibinfo {volume} {106}},\ \bibinfo
  {eid} {053907} (\bibinfo {year} {2009})}\BibitemShut {NoStop}%
\bibitem [{\citenamefont {Evans}\ and\ \citenamefont {Fan}(2014)}]{Evans_3}%
  \BibitemOpen
  \bibfield  {author} {\bibinfo {author} {\bibfnamefont {R.~F.~L.}\
  \bibnamefont {Evans}}\ and\ \bibinfo {author} {\bibfnamefont {W.~J.}\
  \bibnamefont {Fan}},\ }\href@noop {} {\bibfield  {journal} {\bibinfo
  {journal} {Appl. Phys. Lett.}\ }\textbf {\bibinfo {volume} {105}},\ \bibinfo
  {eid} {192405} (\bibinfo {year} {2014})}\BibitemShut {NoStop}%
\bibitem [{\citenamefont {Chantrell}\ and\ \citenamefont
  {Wohlfarth}(1985)}]{Chantrell}%
  \BibitemOpen
  \bibfield  {author} {\bibinfo {author} {\bibfnamefont {R.~W.}\ \bibnamefont
  {Chantrell}}\ and\ \bibinfo {author} {\bibfnamefont {E.~P.}\ \bibnamefont
  {Wohlfarth}},\ }\href@noop {} {\bibfield  {journal} {\bibinfo  {journal}
  {phys. stat. sol.}\ }\textbf {\bibinfo {volume} {91}},\ \bibinfo {pages}
  {619} (\bibinfo {year} {1985})}\BibitemShut {NoStop}%
\bibitem [{\citenamefont {Kazantseva}\ \emph
  {et~al.}(2009{\natexlab{b}})\citenamefont {Kazantseva}, \citenamefont
  {Hinzke}, \citenamefont {Chantrell},\ and\ \citenamefont
  {Nowak}}]{Kazantseva_3}%
  \BibitemOpen
  \bibfield  {author} {\bibinfo {author} {\bibfnamefont {N.}~\bibnamefont
  {Kazantseva}}, \bibinfo {author} {\bibfnamefont {D.}~\bibnamefont {Hinzke}},
  \bibinfo {author} {\bibfnamefont {R.~W.}\ \bibnamefont {Chantrell}}, \ and\
  \bibinfo {author} {\bibfnamefont {U.}~\bibnamefont {Nowak}},\ }\href@noop {}
  {\bibfield  {journal} {\bibinfo  {journal} {Euro. phys. Lett.}\ }\textbf
  {\bibinfo {volume} {86}},\ \bibinfo {pages} {27006} (\bibinfo {year}
  {2009}{\natexlab{b}})}\BibitemShut {NoStop}%
\bibitem [{\citenamefont {Barker}\ \emph
  {et~al.}(2010{\natexlab{b}})\citenamefont {Barker}, \citenamefont {Evans},
  \citenamefont {Chantrell}, \citenamefont {Hinzke},\ and\ \citenamefont
  {Nowak}}]{Barker_2}%
  \BibitemOpen
  \bibfield  {author} {\bibinfo {author} {\bibfnamefont {J.}~\bibnamefont
  {Barker}}, \bibinfo {author} {\bibfnamefont {R.~F.~L.}\ \bibnamefont
  {Evans}}, \bibinfo {author} {\bibfnamefont {R.~W.}\ \bibnamefont
  {Chantrell}}, \bibinfo {author} {\bibfnamefont {D.}~\bibnamefont {Hinzke}}, \
  and\ \bibinfo {author} {\bibfnamefont {U.}~\bibnamefont {Nowak}},\
  }\href@noop {} {\bibfield  {journal} {\bibinfo  {journal} {Appl. Phys.
  Lett.}\ }\textbf {\bibinfo {volume} {97}},\ \bibinfo {eid} {192504} (\bibinfo
  {year} {2010}{\natexlab{b}})}\BibitemShut {NoStop}%
\bibitem [{\citenamefont {Weller}\ and\ \citenamefont {Moser}(1999)}]{Weller}%
  \BibitemOpen
  \bibfield  {author} {\bibinfo {author} {\bibfnamefont {D.}~\bibnamefont
  {Weller}}\ and\ \bibinfo {author} {\bibfnamefont {A.}~\bibnamefont {Moser}},\
  }\href@noop {} {\bibfield  {journal} {\bibinfo  {journal} {IEEE Trans.
  Magn.}\ }\textbf {\bibinfo {volume} {35}},\ \bibinfo {pages} {4423} (\bibinfo
  {year} {1999})}\BibitemShut {NoStop}%
\bibitem [{\citenamefont {Ostler}\ and\ \citenamefont {et.al.}(2012)}]{Ostler}%
  \BibitemOpen
  \bibfield  {author} {\bibinfo {author} {\bibfnamefont {T.}~\bibnamefont
  {Ostler}}\ and\ \bibinfo {author} {\bibnamefont {et.al.}},\ }\href@noop {}
  {\bibfield  {journal} {\bibinfo  {journal} {Nat. Commun}\ }\textbf {\bibinfo
  {volume} {3}},\ \bibinfo {pages} {666} (\bibinfo {year} {2012})}\BibitemShut
  {NoStop}%
\end{thebibliography}

%

\end{document}